\documentclass[12pt, preprint, natbib209]{aastex}
 
 \shorttitle{Enhanced Mass Loss in Cepheids}
 \shortauthors{Neilson \& Lester}
  
\begin{document}

 \title{On the Enhancement of Mass Loss in Cepheids Due to Radial Pulsation}
 \author{Hilding R. Neilson}
\affil{Department of Astronomy \& Astrophysics, University of Toronto, 50 St. George Street, Toronto, ON, Canada M5S 3H4}
\email{neilson@astro.utoronto.ca}
\author{John B. Lester}
\affil{University of Toronto Mississauga, Mississauga, ON, Canada L5L 1C6}

\begin{abstract}
An analytical derivation is presented for computing mass--loss rates of Cepheids by using the method of \cite{Castor1975} modified to include a term for momentum input from pulsation and shocks generated in the atmosphere.  Using this derivation, mass--loss rates of Cepheids are determined as a function of stellar parameters.  When applied to a set of known Cepheids, the calculated mass--loss rates range from $10^{-10}$--$10^{-7}M_\odot/yr$, larger than if the winds were driven by radiation alone.  Infrared excesses based on the predicted mass--loss rates are compared to observations from optical interferometry and IRAS, and predictions are made for Spitzer observations.  The mass--loss rates are consistent with the observations, within the uncertainties of each.  The rate of period change of Cepheids is discussed and shown to relate to mass loss, albeit the dependence is very weak.  There is also a correlation between the large mass--loss rates and the Cepheids with slowest absolute rate of period change due to evolution through the instability strip.  The enhanced mass loss helps illuminate the issue of infrared excess and the mass discrepancy found in Cepheids.
\end{abstract}

\section{Introduction}
Cepheid variable stars are powerful tools for many aspects of astrophysics.  Cepheids are excellent laboratories for stellar physics while also providing a period--luminosity relation that is essential for extragalactic and cosmological studies.  By observing both the intensity variation across the stellar disk as well as the variation of the angular diameter as function of pulsation phase, the emerging field of optical interferometry is extracting further information.  This new information demands more sophisticated Cepheid models of the stellar interior and atmosphere, but also provides a more robust distance to individual Cepheids, helping to calibrate the period--luminosity relation that is more precise than any other calibration. 

\cite{Lane2000} reported the first detection of the radius variation of a Cepheid, $\zeta$ Geminorum, where the authors determined the mean angular diameter and incorporated radial velocity data to derive the distance. \cite{Kervella2004a} observed seven Cepheids, determining the variations of the angular diameter for four of them in the K--band with a precision better than $10\%$.  In turn, these authors were able to calculate distances to these Cepheids with a statistical  uncertainty of only a few percent, more accurate than HIPPARCOS parallaxes. More recently, interferometric K--band observations of Polaris, $\delta$ Cephei, and $l$ Carinae provided evidence of circumstellar shells \citep{Kervella2006, Merand2006}.  These circumstellar shells cause an infrared excess of a few percent and increase the uncertainty of the angular diameter and the variations. The larger uncertainty, in turn, decreases the precision of the distance determination.  The existence of these circumstellar shells pose a problem both for understanding the structure of Cepheids and for the calibration of a more precise period--luminosity relation.

\cite{Merand2007} continued to explore the phenomenon of circumstellar shells with interferometric observations of Y Oph and the non--pulsating yellow supergiant $\alpha$ Per.  They found evidence of a circumstellar shell about Y Oph contributing approximately $5\%$ of the total flux, larger than the excesses from previous observations. The yellow supergiant $ \alpha$ Per, on the other hand, was found to have no flux excess.  This result lends credence to the idea that the shells are related to radial pulsation.

Interferometric observations appear to have raised a new set of questions regarding the evolution of  Cepheids but the detection of circumstellar shells may be a manifestation of an older problem: the idea of mass loss in Cepheids.  Throughout the 1980's, mass loss was argued as being the cause of the mass discrepancy in Cepheids \citep{willson1989, brunish1989}.  The Cepheid mass discrepancy \citep[see][for a detailed explanation]{Cox1980} is the difference between predicted Cepheid masses estimated using evolutionary stellar models and those derived using pulsation models.  The difference, at that time, was as much as 30\% to 40\%.  \cite{willson1989} argued there is significant mass loss related to shocks generated in the atmospheres of Cepheids.  Furthermore the enhanced mass loss would slow the evolution of the Cepheids on the blue loop and increase the mass lost to winds even more.  After some evolution, Cepheids could lose enough mass to reconcile the difference between evolutionary and pulsational predictions of the mass of a Cepheid. Other theories were proposed to explain the discrepancy, such as convective core overshoot in the progenitor models of Cepheids, which could generate a steeper mass--luminosity relation \citep{Keller2006,Keller2008}. Another theory proposes missing opacities in the evolutionary models.  In the end, the introduction of the OPAL opacities \citep{Rogers1992} had the greatest effect on reconciling the mass discrepancy; \cite{Moskalik1992} found the period estimates to be reduced for the same set of stellar parameters ($L$, $M$, $T_{\rm{eff}}$, X, Z) using the OPAL opacities compared to the Los Alamos opacities. This result seemed to end the theoretical study of mass loss in Cepheids as the current mass discrepancy is of the order of $10\%$ for short period Cepheids, and decreasing with longer period \citep{Bono2001}.

While the theoretical study of mass loss in Cepheids is limited, much more work has been presented on the observational side.  \cite{McAlary1986} used IRAS to observe a sample of Cepheids to search for infrared excess.  They found small infrared excesses in twelve Cepheids and estimated mass--loss rates of the order $10^{-9}\rm{-}10^{-8}\rm{M_\odot/yr}$. More IRAS observations \citep{Deasy1988} found upper limits of mass loss rates ranging from $10^{-10}\rm{-}10^{-6} \rm{M_\odot/yr}$.  These mass--loss rates were calculated by assuming the observed flux excess is caused by dust emission.  Radio observations using the VLA set upper limits for mass--loss rates of order $10^{-9}\rm{-}10^{-7}\rm{M_{\odot}/yr}$ for the classical Cepheids FF Aql, $\eta$ Aql, SU Cas, $\delta$ Cep  and T Mon  \citep{Welch1988} by assuming the emission is due to hot ionized gas.   With observations from the IUE satellite of the Cepheid binary SU Muscae, \cite{Rodrigues1992} determined an upper limit for the mass--loss rate $\dot{M} \le 7\times 10^{-10}\rm{M_\odot/yr}$.  \cite{Bohm-Vitense1994} found a mass--loss rate for {\it{l}} Carinae of $2\times 10^{-5}\rm{M_\odot/yr}$ with an uncertainty of two orders of magnitude, again using IUE observations.  Collectively, these observations provide significant evidence for mass loss in Cepheids.  Indirect evidence of mass loss is the detection of circumstellar shells surrounding RS Pup \citep{Havlen1972} and SU Cas \citep{Turner1984} from optical observations, though both Cepheids are associated with reflection nebulae.

The current state of the Cepheid mass discrepancy has been reviewed by \cite{Bono2006}, where four plausible solutions are presented: extra convective mixing, rotation, stellar opacity and mass loss.  If one increases the amount of convective core overshoot in main sequence evolutionary models of Cepheid progenitors then more hydrogen mixes into the core and creates a more massive helium core.  This causes  a more luminous Cepheid at fixed mass.  Increasing the rotation in Cepheid progenitors has  the same effect as extra convective mixing while increasing the radiative stellar opacity would affect the driving of pulsation, causing a longer period for the same mass or conversely a lower mass for the same period of pulsation.  Because new stellar opacities went far in resolving the mass discrepancy before, it is possible the discovery of new radiative opacities at temperatures $10^5$ K to $10^6$ K could still make a difference.  Mass loss is also a possibility, and the connection between mass loss and pulsation has been explored recently for other types of stars, such as luminous blue variables \citep{Guzik2005}, asymptotic giant branch stars \citep{Hofner1999} and Miras \citep{Hofner1997}. 

If circumstellar shells surrounding Cepheids are generated by mass loss then the observations of \cite{Merand2007} imply mass loss is driven by pulsation, however, this does not preclude other driving mechanisms. The most common mechanisms for driving a stellar wind are: dust driving, radiative line--driving, coronal winds, and magnetic fields \citep{Lamers1999}.  Dust driving is important in stars where dust can form in the atmosphere.  This leads to an increase of the opacity of the star and radiation pressure accelerates the dust away from the star.  If the dust is ionized then the particles have a larger effective collisional cross--section and causes more material to be ejected.  Since dust must form in the atmosphere for this to be important, the star's effective temperature must be cooler than $2000$ K; Cepheids have effective temperatures of order $4000$ K -- $6500$ K.  Dust is not a likely factor though it can form some distance from the star where the temperature of the gas is much less when the gas is optically thin.  

Radiative line--driving has been explored thoroughly for most stars and is considered to be most important for hotter stars where carbon, nitrogen and oxygen atoms are ionized in the atmosphere causing a larger effective collisional cross--section.  The ions are accelerated by radiation and interact with other atoms in the atmosphere, in turn accelerating them outwards.  For stars with an effective temperature of order $6000$ K, neon, iron, hydrogen and helium may play the role that carbon, nitrogen and oxygen play in hotter stars \citep{Abbott1982}.  Therefore radiative line--driving must be considered in any analysis of mass loss in Cepheids. 

 To have a coronal wind, a star must have a corona where the temperature is very large.  \cite{Sasselov1994} and \cite{Schmidt1982, Schmidt1984b, Schmidt1984a} found only weak chromospheres for Cepheids implying there are no hot coronae;  coronal driving is an unlikely source. 
 
 The presence of magnetic fields is uncertain and controversial in Cepheids. \cite{Plachinda2000} detected a $100$ $\rm{G}$ magnetic field in $\eta$ Aquilae but \cite{Wade2002} could not verify that result, inferring an upper limit a $10$ $\rm{G}$ field.  However, Polaris has recently been observed to be a soft X--ray source, though this may be due to its binary companion; far UV observations of Polaris imply the possible existence of warm winds, shocks or magnetic activity \citep{Engle2006}.  Magnetic fields seem an unlikely cause of winds in Cepheids.  Thus one needs to be concerned with only one of those four possibilities: radiative line driving.

The purpose of this study is to explore the role radial pulsation plays in driving mass loss in classical Cepheids when coupled with radiative line--driving. \cite*{Castor1975}, hereafter CAK, devised a method to solve for the radiation--driven wind structure of a star. Here that method is modified to include a simple function describing additional momentum input into the wind.   The second section describes the derivation of the CAK momentum equation for a pulsating star, including pulsation and shock physics. Section three explores the mass--loss rates for observed Cepheids, and Section 4 compares predicted flux excesses with those determined by \cite{Deasy1988}, \cite{Kervella2006} and \cite{Merand2006, Merand2007}. The fifth section will consider the dependence of the rate of period change, $\dot{P}/P$, on mass loss.  

\section{Pulsation Enhanced CAK Method}
\cite{Castor1975} proposed an isothermal wind model to describe the mass--loss rates and wind structures of stars.  The method is a powerful, yet simple, analytical tool for understanding the effects of radiative line driving because it requires knowledge of only the mass, radius and luminosity of the star. The authors assume the forces due to the radiative lines and the pressure gradient are functions of the local velocity gradient; this results in having the conservation of momentum written in a form that can be solved for the velocity as a function of distance from the stellar surface.   The CAK wind model is the standard tool for understanding winds in O, B, and A stars \citep{Owocki2005, Watanabe2006} and possibly F and G supergiants \citep{Lamers1999}.  A review of the analytic derivation can be found in \cite{Lamers1999}.

\subsection{Derivation for Pulsation Driven Winds}
One can derive the equations governing the mass--loss rate for a radially pulsating star by following the derivation given by \cite{Lamers1999}, who start with the momentum equation for the wind including the effects of continuum and radiative line driving, gravity and gas pressure,
\begin{equation}\label{e1}
v\frac{dv}{dr} = - \frac{GM_*}{r^2} + \frac{1}{\rho}\frac{dp}{dr} + g_e + g_L.
 \end{equation}
The functions $g_e$ and $g_L$ describe the radiation force per unit mass due to electron scattering and line opacity respectively.  The expression for $g_e$ is
\begin{equation}\label{e2}
g_e = \frac{\sigma_e L_*}{4\pi r^2 c} = \frac{GM_*}{r^2} \Gamma_e
\end{equation}
where $\Gamma_e \equiv \sigma_e L_*/(4\pi c GM_*) $, and $\sigma_e = \sigma_T N(e)$ is the electron scattering opacity as a function of the Thomson scattering opacity and the number of electrons.  The electron scattering term is clearly dependent on the composition and temperature of the star, but for this analysis is assumed not to vary significantly.  The acceleration due to lines is found by using the Sobolov approximation; the final form is
\begin{equation}\label{e3}
g_L = Cr^{-2}\left(r^2v\frac{dv}{dr}\right)^\alpha ,
\end{equation}
where
\begin{equation}\label{e4}
C = \frac{\sigma_e L_* k}{4\pi c}\frac{Z}{Z_\odot}\left[\frac{\sigma_e v_{\rm{th}} \dot{M}}{4\pi}\right]^{-\alpha} \left[ \frac{10^{-11} n_e}{W}\right]^{\delta}.
\end{equation}
The function $C$ is treated as a constant, although this is not strictly true.  The parameters $\alpha$, $k$ and $\delta$ are force multiplier parameters calculated using model atmospheres to measure the radiative acceleration due to lines.  The values of $\alpha$ and $\delta$ tend to be of the order $0.5$ and $0.1$ respectively, while $k$ can vary significantly: for stars $T_{\rm{eff}} < 10^4$ K $k \approx 0.1$ \citep{Abbott1982, Shimada1994, Pauldrach1986}.  The ratio of the electron density with the geometric dilution factor $W(r) = (1-\sqrt{1-R_*/r})/2$ can vary for all stars implying $C$ is not constant, but the ratio has the exponent $\delta$ that is small.  Therefore the term to the power $\delta$ is ignored by assuming it is approximately unity.  It should be noted, however, in Cepheids the ionization rate of hydrogen and helium is dependent on the phase of pulsation.  The number density of electrons could thus vary as a function of phase and play a small role.  If the number density of electrons is increased then the mass--loss rate  that is calculated may be larger as a result, but not by more than an order of magnitude because the term $\Gamma_e $ is much less than unity.  Also $C$ has an explicit and implicit metallicity dependence found from calculations on the contributions to radiative line driving for various ions as a function of gravity and effective temperature \citep{Abbott1982, Shimada1994}.

The pressure term is rewritten in terms of the velocity by assuming the wind is isothermal
\begin{equation}\label{e5}
\frac{1}{\rho} \frac{dp}{dr} = -\frac{2a^2}{v}\frac{dv}{dr}- \frac{2a^2}{r},
\end{equation}
where $a$ is the isothermal sound speed.  The resulting total form for the conservation of momentum is
\begin{equation}\label{e6}
v\frac{dv}{dr} = - \frac{GM_*(1-\Gamma_e)}{r^2} + \frac{a^2}{v}\frac{dv}{dr}  + \frac{2a^2}{r} + Cr^{-2}\left(r^2v\frac{dv}{dr}\right)^\alpha.
\end{equation}
Equation \ref{e6} does not include the effect of pulsation. Radial pulsation injects momentum into the wind due to the acceleration of the outer layers of the atmosphere of the star.  Radial pulsation also generates shocks in the interior that propagate to the surface, depositing energy into the wind. This means there are two additional sources of acceleration in the atmosphere of the Cepheid.  For this derivation, it is assumed the accelerations due to pulsation and shocks are determined at the surface of the star so that they are only a function of time and global parameters.  In the wind, these additional acceleration terms are modified by a dissipation factor, $(r/R_*)^{-\nu}$ where $\nu > 0$. Therefore the sum of the acceleration due to pulsation and the acceleration due to shocks are written as $\zeta(r/R_*)^{-\nu}$. The function $\zeta$ is the acceleration at the surface of the star due to the sum of the pulsation and shocks that is written in terms of the global quantities of the star. The function is derived in detail in the next section.  The full equation for the conservation of momentum including pulsation is
\begin{equation}\label{e7}
r v\frac{dv}{dr} = - \frac{GM_*(1-\Gamma_e)}{r^2} + \frac{a^2}{v}\frac{dv}{dr} + \frac{2a^2}{r}  + Cr^{-2}\left(r^2v\frac{dv}{dr}\right)^\alpha + \zeta \left(\frac{r}{R_*}\right)^{-\nu}.
\end{equation}
Rearranging terms and simplifying, Equation \ref{e7} becomes
\begin{equation}\label{e8}
C\left(r^2v\frac{dv}{dr}\right)^\alpha - \left(1 - \frac{a^2}{v^2}\right)r^2v\frac{dv}{dr}  -GM_*(1-\Gamma_e)  + 2a^2 r + \zeta R_*^2  \left(\frac{r}{R_*}\right)^{(2-\nu)} = 0.
\end{equation}
Equation \ref{e8} represents the quasi--static approximation of the momentum equation for the wind of a pulsating star.  The differential equation can be solved in a series of snapshots where the velocity calculated is based on the conditions of the star at the given time, however this does not change the mass--loss rate at that instant.  The mass--loss rate depends on the velocity and the density of the wind as it is ejected, but the wind does not feed back onto the layers that generate the wind.  Therefore the time averaged mass--loss rate can be determined by this quasi--static approximation but the velocity of the wind will be unphysical near the surface of the star.  Hence only the change of the luminosity and radius and other quantities dependent on them are considered in the problem.

Because the implicit effects of time are ignored, the differential equation can be solved in the same manner as the non--pulsating form.  From Figure 4 of \cite{Cassinelli1979}, it is clear there is only one solution to the non--pulsating differential equation, the solution where the velocity as a function of distance from the star satisfies both regularity and singularity conditions at some critical point, $r_c$.  It must be noted the quantity $r_c$ is a mathematical construct for solving the differential equation and is not necessarily a physical quantity.  If $F = 0$ represents the momentum equation, Equation \ref{e8}, then
\begin{eqnarray}\label{e9}
&&\left(\frac{\partial F}{\partial v^\prime}\right)_c = 0 \mbox{\hspace{2.5cm}Singularity Condition} \\
&&\label{e10} \left(\frac{\partial F}{\partial r}\right)_c + \left(v^\prime \frac{\partial F}{\partial v}\right)_c = 0 \mbox{\hspace{0.5cm}Regularity Condition},
\end{eqnarray}
where $\prime$ denotes the derivative with respect to $r$.  While it may seem that using the velocity structure to determine a valid solution is contradictory with previous statements, this would be a reasonable approximation if the star were non--pulsating.  The amount of mass ejected at time $t$ is not be affected by the amount of mass ejected an instant later.  At that instant when the mass is ejected it has a velocity structure for a steady flow that provides a satisfactory solution to the differential equation, though this ignores the effect of clumping.  The singular and regularity conditions become
\begin{equation}\label{e11}
\left(1 - \frac{a^2}{v_c^2}\right) r_c^2 v_c v_c^\prime = \alpha C\left(r_c^2v_cv_c^\prime\right)^\alpha ,
\end{equation}
\begin{eqnarray}\label{e12}
&&\nonumber \frac{2\alpha C}{r_c}\left(r_c^2v_cv_c^\prime\right)^{\alpha} - \left(1 -  \frac{a^2}{v_c^2}\right)2r_c v_c v_c^\prime + 2a^2 +\zeta R_* \left(\frac{r_c}{R_*}\right)^{1-\nu} \\
&& + \frac{v_c^\prime}{v_c}\left[ \alpha C(r_c^2 v_c v_c^\prime)^\alpha - \left(1 - \frac{a^2}{v_c^2}\right) r_c^2 v_c v_c^\prime - 2\frac{a^2}{v_c^2}r_c^2v_cv_c^\prime\right] =  0
\end{eqnarray}
respectively, where one should recall  $d\zeta/dr =0$ as $\zeta$ is defined at the surface of the star. 

One may now use the momentum equation and the singularity and regularity conditions to determine the velocity and velocity gradient at the critical point by first substituting the momentum equation into the singularity condition, and eliminating the quantity $C$,
\begin{equation}\label{e13}
 \mbox{\hspace{-0.5cm}}r^2_cv_cv_c^\prime = \left(\frac{\alpha}{1-\alpha}\right)\left(1-\frac{a^2}{v_c}\right)^{-1}\left[GM_*(1-\Gamma_e) -2a^2r_c - \zeta R_*^2 \left(\frac{r_c}{R_*}\right)^{(2-\nu)}\right].
\end{equation}
Combining the singularity condition with the regularity condition will determine the velocity gradient at the critical point and using that result with Equation \ref{e13} one finds the velocity at the critical point
\begin{equation}\label{e14}
v_c^\prime = \frac{v_c}{r_c}\left[1 + \frac{\zeta R_*(2-\nu)}{2a^2} \left(\frac{r_c}{R_*}\right)^{(1-\nu)}\right]^{1/2},
\end{equation}
\begin{eqnarray}\label{e15}
&&v_c^2 = a^2  + \left[1 + \frac{\zeta R_*  (2-\nu)}{2a^2}\left(\frac{r_c}{R_*}\right)^{1-\nu}\right]^{-1/2} \left(\frac{\alpha}{1-\alpha}\right) \\
&& \times\left[\frac{GM_*(1-\Gamma_e)}{r_c} - 2a^2 - \zeta R_*\left(\frac{r_c}{R_*}\right)^{1-\nu}\right].
\end{eqnarray}
These two expressions give the velocity and velocity gradient at the critical point and form a starting point for solving the momentum equation.  The expression $v_c^\prime$ is combined with the singularity condition to determine the radiative driving constant $C$ in terms of the radius and velocity at the critical point and various terms from the momentum equation
\begin{eqnarray}\label{e16}
&&\nonumber C = \left(\frac{1}{1-\alpha}\right) \left[1 + \frac{\zeta R_*(2-\nu) }{2a^2} \left(\frac{r}{R_*}\right)^{(1-\nu)}\right]^{-\alpha/2} \frac{1}{r_c^\alpha v_c^{2\alpha}} \\ 
&& \times\left[GM_*(1-\Gamma_e) - 2a^2r_c - \zeta R_*^2\left(\frac{r_c}{R_*}\right)^{2-\nu}\right].
\end{eqnarray}
Equating the two expressions for $C$, Equation \ref{e4}  and \ref{e16}, and using the function for the critical velocity, Equation \ref{e15}, yields the relation for the mass--loss rate 
\begin{eqnarray}\label{e17}
&&\mbox{\hspace{-0.75cm}}\nonumber\dot{M} = \left[\frac{\sigma_e L_* k }{4\pi c}\frac{Z}{Z_\odot}(1-\alpha)\right]^{1/\alpha}\left(\frac{4\pi}{\sigma_e v_{th}}\right)  \left[GM_*(1-\Gamma_e) - 2a^2r_c - \zeta R_*^2\left(\frac{r_c}{R_*}\right)^{2-\nu}\right]^{-1/\alpha}\\
&&\mbox{\hspace{-0.75cm}}\nonumber \times \left\{a^2r_c\left[1 + \frac{\zeta R_*(2-\nu)}{2a^2} \left(\frac{r_c}{R_*}\right)^{(1-\nu)}\right]^{1/2}\right. \\
&&\mbox{\hspace{-0.75cm}}\left. +\left(\frac{\alpha}{1-\alpha}\right)\left[GM_*(1-\Gamma_e) - 2a^2r_c - \zeta R_*^2\left(\frac{r_c}{R_*}\right)^{2-\nu}\right]\right\}.
\end{eqnarray}
It is now shown the mass--loss rate of a Cepheid can be calculated by knowing the global parameters for the star and the critical point.  \cite{Lamers1999} argue that if one assumes a value for $r_c$ then the velocity structure can be solved and the density structure can be derived from the velocity.  The continuum optical depth of the wind is calculated from the density.  This lends itself to a condition for the solution of the differential equation.  The stellar surface is generally defined where the mean optical depth equals $2/3$, therefore the optical depth of the wind, given by the integration of the density and opacity from infinity to the stellar surface, must be $2/3$.  The inclusion of pulsation and shocks, however, means it is not so clear this criterion needs to be satisfied for a wind to be driven.   Because there is no ideal criterion to use that will  give a clear solution, the optical depth $\tau = 2/3$ will be used. The mass--loss rate is tested and shown to verify that it varies only weakly for different values of the critical point; hence it is reasonable to use the integral of the optical depth as a criterion for solving the momentum equation. Therefore at some phase of pulsation, one can calculate the instantaneous mass loss rate by adopting a value for $r_c$ and solving the wind structure and comparing the velocity predicted at the surface with the velocity due to pulsation;  a solution is reached when the two values match.

\subsection{Defining the Acceleration Due to Pulsation and Schocks}

In deriving the solution for the enhanced mass loss due to pulsation and shocks, it was assumed the function $\zeta$, which represents the sum of the acceleration from both pulsation  and shocks, depends only on the global parameters that describe the Cepheid, such as effective temperature, radius, amplitude of radius variation and period of pulsation.  In this subsection, the formulation for the acceleration will be defined and justified.

For simplicity it is advantageous to consider a one--zone model for the radial pulsation of the Cepheid with period $P$.  In that case the radius of the Cepheid as a function of time is  written as $\Delta R(t) = -\Delta R \cos(\omega t)$, where $\Delta R$ is the amplitude of pulsation.  The amplitudes of the nearest Cepheids are readily available from radial velocity studies, and a list has been compiled by \cite{Moskalik2005}.  The minus sign in the equation defines a phase of zero to correspond to the minimum radius and $\omega = 2\pi/P$ is  the angular frequency of pulsation. The acceleration of the surface of the Cepheid due to pulsation is thus 
\begin{equation}\label{e18}
a_{\rm{puls}} = d^2[\Delta R(t)]/dt^2 = \omega^2 \Delta R \cos(\omega t).
\end{equation}

The luminosity varies as a function of time as well.  Since there is a well--known phase lag, such that the maximum luminosity occurs roughly a quarter period before the maximum radius, the quasi--adiabatic approximation is used to describe the luminosity such that $\Delta L(t) = \Delta L\sin(\omega t)$.  The electron scattering opacity can be written via Kramer's law to be $\sigma_e \propto \rho T^{-3.5}$, implying it too varies due to pulsation because both the temperature and density vary as a function of time.  Because the temperature goes as $T_{\rm{eff}} \propto [L_*(t)/(R_*(t)^2]^{1/4}$, the variation of the temperature is
\begin{equation}\label{e19}
\frac{\Delta T_{\rm{eff}}(t)}{\bar{T}_{\rm{eff}}} =\frac{1}{4} \frac{\Delta L_*(t)}{\bar{L}_*} - \frac{1}{2}\frac{\Delta R_*(t)}{\bar{R}_*}.
\end{equation}
In addition, the linear perturbation of the density in the one zone model is
\begin{equation}\label{e20}
\frac{\Delta \rho (t)}{\bar{\rho}} = - 3\frac{\Delta R_*(t)}{\bar{R}_*},
\end{equation}
meaning the variation of the electron scattering opacity is
\begin{equation}\label{e21}
\frac{\Delta \sigma_e}{\bar{\sigma_e}} = -\frac{5}{4}\frac{\Delta R}{\bar{R}_*} - \frac{7}{8}\frac{\Delta L}{\bar{L}_*},
\end{equation}
where variables denoted with a bar are the mean values of the variable over a pulsation period. Also, because the effective temperature will vary as a function of phase then so will the isothermal sound speed, which goes as $a \propto T^{1/2}$.  This defines all of the quantities that are assumed to vary due to pulsation.

The role of shocks is not as easily quantified as there is no simple method to describe the acceleration of gas by spherically symmetric shocks.  \cite{Willson1976} and \cite{Willson1979} develop a model for a shock in the atmosphere of long period variables where the shock is periodic with the pulsation. However \cite{Mathias2006} and \cite{Sasselov1990}, for example, find evidence for multiple shocks per period. Also, while the mechanism for generating shocks is understood to be related to both the opacity mechanism and the $\gamma$--mechanism, the strength of shocks is uncertain. An analysis of shock behavior in $\delta$ Cephei was conducted by \cite{Fokin1996},  who modeled the atmosphere of  $\delta$ Cephei and traced the difference between the post--shock velocity of the gas and the pre--shock velocity of the gas in the frame of reference of the traveling shock as a function of the phase of pulsation. Our work uses that information to estimate the acceleration of the gas in Cepheids due to shocks. Here it is assumed that the velocity of the shock itself goes to zero at the surface of the Cepheid.  This assumption is justified based on evidence that Cepheids have weak chromospheres generated by shocks \citep{Sasselov1994}.  Therefore the difference between the velocity of the pre-- and post--shocked gas at the stellar surface in the frame of the shock is the same as in the rest frame.  It is also assumed that the velocities in \cite{Fokin1996} are valid in a one zone model, i.e. these velocities are at the surface of the Cepheid.  

From \cite{Matzner1999} and \cite{Klimishin1981}, one can assume the pressure of the post--shocked gas is proportional to the mean energy density
\begin{equation}\label{e22}
u^2 = \frac{E}{m},
\end{equation}
where $u,E, m$ represent the velocity, mean energy density and the mass of the post shock gas, respectively.  The mean energy density is roughly constant, while the mass is not.  For a central explosion, such as a supernova, one can write the mass as $ m \approx \rho r^3$, where $r$ is the scale of the post shock gas from the shock. However, shocks in Cepheids are generated relatively near the surface, therefore we can write the mass as $m \approx \rho R_*^2 \Delta r$, where $\Delta r$ is the thickness of the shock front. Thus the velocity relative to the sound speed inside the star, $c_s$, scales as
\begin{equation}\label{e23}
\frac{u}{c_s} = \Omega \left(\frac{m}{M_*}\right)^{-1/2} = \Omega \left(\frac{\rho 4\pi R_*^2 \Delta r}{\bar{\rho}4\pi R_*^3/3}\right)^{-1/2},
\end{equation}
where $\Omega$ is a constant, and $\Delta r$ is the mean free path times the pre--shock pressure divided by the change of pressure \citep{ZelDovich1967},
\begin{equation}\label{e24a}
\Delta r = l \frac{p}{\Delta p}.
\end{equation}
The pre--shock pressure is $c_s^2\rho$ at the surface and the change of pressure is the sound speed at the radius where the shock is formed times the change of density.  The shock is formed at the partial ionization zone, which has the same temperature for all Cepheids.  The mean free path is $1/\kappa \rho$, and the opacity of the shock is also constant.  Therefore the thickness of the shock front is 
\begin{equation}\label{e25a}
 \Delta r = \frac{1}{\kappa \Delta \rho} \frac{T_{\rm{eff}}}{T_{\rm{ionization}}} =  \frac{1}{\kappa \Delta \rho} \frac{L_*^{1/4}R_*^{-1/2}}{L^{1/4}(r)r^{-1/2}} = \frac{1}{\kappa \Delta \rho} \frac{r^{1/2}}{R_*^{1/2}} ,
\end{equation}
where the luminosity is the same at both $r$ and $R_*$, and the radius where the shock is formed is roughly constant for all Cepheids. Thus the speed of the gas due to a shock, given by Equation \ref{e23}, is proportional to 
\begin{equation}\label{26a}
u \propto c_s\left(\bar{\rho}R_*^{3/2}\right)^{1/2}.
\end{equation}
Therefore the ratio of the velocity of gas for any Cepheid relative to the prototypical Cepheid, $\delta$ Cephei, is
\begin{equation}\label{e26}
\frac{u}{u_{\delta}} = \frac{c_s \bar{\rho}^{1/2}R_*^{3/4}}{c_{s,\delta}\bar{\rho}_{\delta}^{1/2}R_{\delta}^{3/4}},
\end{equation}
where $\delta$ refers to $\delta$ Cep. The sound speed is proportional to the square root of the effective temperature which, in turn, is proportional to the luminosity and the radius, $T_{\rm{eff}} \propto L^{1/4}R_*^{-1/2}$.  The relative shock speed is now $u/u_\delta = (L*^{1/8}\bar{\rho}R_*^{1/2})/(L_\delta^{1/8}\bar{\rho}_\delta R_\delta^{1/2})$.  The radius can be expressed in terms of the mean density and the mass, $R_* \propto \bar{\rho}^{-1/3}M^{1/3}$ while the mean density is a function of the period of pulsation $\bar{\rho} = (Q/P_*)^2$ where $Q$ is the pulsation constant. \cite{Fernie1967} found that pulsation constant is not strictly constant, and actually varies as $Q \propto P^{1/8}$. Equation \ref{e26} is rewritten in terms of the luminosity, period and mass,
\begin{equation}\label{e27}
\frac{u}{u_{\delta}} =\left( \frac{L_*}{L_\delta}\right)^{1/8}\left(\frac{P_*}{P_\delta}\right)^{-7/24}\left(\frac{M_*}{M_\delta}\right)^{5/12}.
\end{equation}
This defines the velocity of the gas due to shocks at the surface of a Cepheid.  

The acceleration of the gas is $a_{\rm{shock}} = P^{-1}du/d\phi $.  In terms of Equation \ref{e27},
\begin{equation}\label{e28}
 a_{\rm{shock}}  =  \frac{P^{-1}}{d\phi}  du_\delta \left( \frac{L_*}{L_\delta}\right)^{1/8}\left(\frac{P_*}{P_\delta}\right)^{-7/24}\left(\frac{M_*}{M_\delta}\right)^{5/12},
\end{equation}
where the quantities describing $\delta$ Cep are taken from the model computed by \cite{Fokin1996}.  The values of $du/d\phi$ are listed in Table \ref{tab1}, where zero phase corresponds to minimum radius.
\begin{table}[t]
\begin{center}
\begin{tabular}{lc}
\hline
Range of Phase & du/d$\phi$ (km/s/phase) \\
\hline
0 - 0.08 & 187 \\
0.1 - 0.12 & 208 \\
0.13 - 0.22 & 120 \\
0.65 - 0.68 & 75 \\
0.73 - 0.76 & 200 \\
0.79 - 0.83 & 240 \\
0.93 - 1 & 187 \\
\hline
\end{tabular}
\end{center}
\caption{The change of speed of shocked gas at the surface of $\delta$ Cephei with respect to phase of pulsation at different phases.  At any phase not within the given ranges the change of speed is zero.  The data are taken from Figure 4, \cite{Fokin1996}.} 
\label{tab1}
\end{table}

Having defined the acceleration of the gas due to shocks and pulsation, the function $\zeta$ is
\begin{equation} \label{e29}
\zeta = \Delta R \omega^2 \cos(\omega t) + \frac{P^{-1}}{d\phi}  du_\delta   \left( \frac{L_*}{L_\delta}\right)^{1/8} \left(\frac{P_*}{P_\delta}\right)^{-7/24}\left(\frac{M_*}{M_\delta}\right)^{5/12},
\end{equation}
where $\zeta$ is dependent only on global parameters: the amplitude of radius variation, luminosity, period, and mass.

\subsection{Solution Space of the Wind Equation}
It is important to explore the momentum equation's parameter space before applying it to real stars.  There may be situations where the combination of parameters do not give a physical solution, which is true for the momentum equation for static stars. There are two regions in the velocity--distance parameter space where there are no solutions: one region is defined where the velocity is subsonic when $GM_*(1-\Gamma_e) - 2a^2r < 0$, the other is when the velocity is supersonic too close to the surface of the star where $GM_*(1- \Gamma_e) - 2a^2r >> 0$.  \cite{Cassinelli1979} reviewed the solution space of a radiative wind.  The first case is explored by analyzing how the addition of pulsation affects the Parker point, $r_p$ \citep{Parker1958}, which is the distance from the star where the effective escape velocity equals the isothermal sound speed.  In the pulsating case the Parker point is defined as
\begin{equation}\label{e30}
GM_*(1- \Gamma_e) - 2a^2 r_p - \zeta R_*^2\left(\frac{r_p}{R_*}\right)^{2-\nu} = 0.
\end{equation}
Second, the boundary where the wind becomes supersonic too close to the surface will be explored by analyzing the parameter space of the critical velocity and the derivative of the critical velocity.

The Parker point is an important quantity in the radiative driving momentum equation as the wind solution must be supersonic at a distance less than
\begin{equation}\label{e31}
r < r_p = \frac{GM_*(1-\Gamma_e)}{2a^2},
\end{equation}
otherwise the velocity gradient $dv/dr <0$ and the $v(r)$ would go to zero.  The inclusion of pulsation adds an extra complication since the Parker point is now defined as the solution to a polynomial equation.  In the momentum equation given by Equation \ref{e8}, it is necessary that
\begin{equation}\label{e32}
F(r) = GM_*(1-\Gamma_e)  - 2a^2 r - \zeta R_*^{\nu} r^{2-\nu} > 0
\end{equation}
when $a^2/v^2 > 1$.  Because this function is non--linear, the Parker point is not a complete definition of the instability, especially since it is mathematically plausible to have a solution where $r_p$ is less than the stellar radius $R_*$.  The true Parker point can be defined by using $F(r_p) = 0$ and $dF(r_p)/dr < 0$.  The second definition ensures $F(r)$ is decreasing before the Parker point.  This is seen if different values of $\nu$ are assumed; consider $\nu = 0,2,3$, which are the quadratic and linear cases for Equation \ref{e30} for the Parker point. In the linear case of $\nu = 2$ the Parker point is
\begin{equation}\label{e33}
r_p = \frac{ GM_*(1-\Gamma_e) - \zeta  R_*^{2}}{2a^2}.
\end{equation}
In this case the Parker point is modified from the traditional definition to one that oscillates about $GM_*(1-\Gamma_e)/2a^2$.  This case does not suffer from multiple solutions, but the cases $\nu = 0, 3$ have two solutions each, where
\begin{equation}\label{e34}
\mbox{for }\nu = 0, \mbox{\hspace{3cm}}r_p = \frac{2a^2 \pm \sqrt{4a^4 + 4\zeta GM_*(1-\Gamma_e)}}{-2\zeta}
\end{equation}
and
\begin{equation}\label{e35}
 \mbox{for }\nu = 3, \mbox{\hspace{1.5cm}}r_p = \frac{GM_*(1-\Gamma_e)}{4a^2} \pm \frac{\sqrt{[GM_*(1-\Gamma_e)]^2 - 8a^2\zeta R_*^{3}}}{-4a^2}
\end{equation}
respectively.  These solutions highlight another potential pitfall,  the term in the square root may be imaginary depending on the phase of pulsation.  In the case of $\nu = 0$, if the phase $0.25\le \phi \le 0.75$ then it is possible for $\zeta < 0$ when the shock acceleration is zero and the acceleration due to pulsation is less than zero. For $\nu = 3$ the Parker point can be imaginary when the function $\zeta$ is large as is the case for short period Cepheids or hotter evolved Cepheids.   If the Parker radius is imaginary then there may not exist a wind solution at that time as the gravity is less than the outward acceleration implying an instability in the star.

Applying the second criterion for a proper solution for the Parker point, one finds the derivative of Equation \ref{e30}
\begin{equation}\label{e35a}
\frac{dF(r_p)}{dr} = -2a^2 - (2-\nu)\zeta R_* \left(\frac{r_p}{R_*}\right)^{1-\nu}.
\end{equation}
Testing this for the case where $\nu = 2$ the derivative is always less than zero. When $\nu = 0$, the Parker point is given by Equation \ref{e34} and substituting this result to solve for $dF(r_p)/dr$
\begin{equation}\label{e35b}
\frac{dF(r_p)}{dr} = \pm\sqrt{a^4 + \zeta GM_*(1-\Gamma_e)}
\end{equation}
implying only the negative square root in the solution for the Parker point as given in \ref{e35} is the true Parker point.  From Equation \ref{e35}, it is clear the Parker point given by the positive square root is less than $R_*$ for most values of $\zeta$.

The second region of interest is based on the wind becoming supersonic too close to the surface of the star, where the boundary is defined as the curve that is tangential to the critical velocity.  Therefore Equation \ref{e15} can be used to analyze any instabilities. There is the obvious possible instability given by the term $\{1 +[ \zeta (2-\nu)R_*/2a^2 ](r_c/R_*)^{1-\nu}\}^{-1/2}$. This term can be imaginary depending on the chosen value of $\nu$ and the phase of pulsation.  If this is imaginary, then both the critical velocity and the derivative of the critical velocity is complex and the will be no real wind solution.  It should be noted that this instability disappears in the case of $\nu = 2$, and  it is possible to compute a mass--loss rate at any phase of pulsation.

The conclusion is that there exists time--dependent perturbations to the Parker point and the critical velocities causing no real wind solutions.  This is different from the non--pulsating CAK method which does not contain any explicit pitfalls.

\subsection{The Power--Law Dependence of the Pulsation and Shock Acceleration}
It is important to understand the role the exponent $\nu$ plays in the pulsation term of the wind equation.  The power law is chosen to represent the dissipation of energy imparted by pulsation as particles in the wind are accelerated outwards.  The use of a power law is an approximation of the detailed physics of the dissipation process.  Without having a firm physical determination of $\nu$, it is necessary to determine how sensitive the total mass--loss rate is to an adopted value of $\nu$.  One way to test the effect of $\nu$ is to derive the change of instantaneous mass--loss rate with respect to $\nu$. This will be done by calculating an approximation of the instantaneous mass--loss rate and solving it for various values of $\nu$.

The sensitivity of the instantaneous  mass--loss rate with respect to $\nu$ can be measured by the quantity $d\ln\dot{M}/d\nu$. The mass--loss rate is given by Equation \ref{e17}, which can be rewritten in the form $\dot{M} = K_1A_1^{-1/\alpha}(A_2 + K_2A_1)$; the term $K_1$  and $K_2 = \alpha/(1-\alpha)$ are constant multipliers, and $A_1$ and $A_2$ are the respective functions of $\nu$. The derivative of the mass--loss rate is
\begin{equation}\label{e36}
\mbox{\hspace{-0.23cm}}\frac{d\ln\dot{M}}{d\nu} = -\frac{1}{\alpha}\frac{1}{A_1}\frac{dA_1}{d\nu} + \frac{dA_2/d\nu + K_2 dA_1/d\nu}{A_2 +K_2 A_1}
\end{equation}
where
\begin{equation}\label{e36a}
A_1 = GM_*(1-\Gamma_e) - 2a^2r_c - \zeta R_*^2\left(\frac{r_c}{R_*}\right)^{2-\nu},
\end{equation}
\begin{equation}\label{e36b}
A_2 = a^2r_c\left[1 + \frac{\zeta R_*(2-\nu)}{2a^2}\left(\frac{r_c}{R_*}\right)^{2-\nu}\right]^{1/2}.+
\end{equation}
This can be considered one term at a time; starting with $A_1$
\begin{equation}\label{e37}
\frac{dA_1}{d\nu} = \zeta  R_* \left(\frac{r_c}{R_*}\right)^{1-\nu} \ln\left(\frac{r_c}{R_*}\right)
\end{equation}
and 
\begin{equation}\label{e38}
 \frac{1}{\alpha}\frac{1}{A_1}\frac{dA_1}{d\nu} = \frac{\ln(r_c/R_*)}{\alpha} \left[ -1 + \frac{GM_*(1-\Gamma_e)/r_c - 2a^2}{\zeta R_* \left(r_c/R_*\right)^{1-\nu}}\right]^{-1}.
\end{equation}
The derivative of the second term is
\begin{equation}\label{e39}
\mbox{\hspace{-0.2cm}} \frac{dA_2}{d\nu} = -\frac{a^2}{2} \left[1 + \frac{\zeta (2-\nu)R_*}{2a^2}\left(\frac{r_c}{R_*}\right)^{(1-\nu)}\right]^{-1/2} \frac{\zeta R_* }{2a^2}\left(\frac{r_c}{R_*}\right)^{(1-\nu)}\left[1+(2-\nu)\ln\left(\frac{r_c}{R_*}\right)\right].
\end{equation}
Starting with the result for $dA_1/d\nu$, it can be seen that there is a large change with respect to $\nu$ if the pulsation term is of order the effective potential, where the enthalpy term given by the isothermal sound speed is relatively insignificant. In the limit the pulsation term is equal to the effective potential then the ratio approaches one and the term in square brackets goes to zero.  Therefore $dA_1/d\nu \rightarrow -\infty$ and the derivative of the mass--loss rate will approach infinity.  The terms $(dA_2/d\nu)/(A_2+K_2A_1)$ and $K_2(dA_1/d\nu)/(A_2+K_2A_1)$ can be shown to have a similar behavior.  Therefore $\nu$ can have a significant effect on the mass--loss rate depending on the combination of parameters.

While the choice of $\nu$ affects the mass--loss rate, it also affects the predicted density structure that is used to calculate the continuum optical depth as a criterion for a solution to the momentum equation.  Consider the mass loss for a pulsating star from two different dissipation laws given by $\nu_1$ and $\nu_2$ where $\nu_2 > \nu_1$. For the power--law $\nu_1$ there is a solution given by a corresponding critical point $r_{c,1}$, which satisfies the Singularity and Regularity condition, Equations \ref{e9} and \ref{e10} respectively, as well as predicting a density structure such that the continuum optical depth is $2/3$.  Using the critical point, $r_{c,1}$, to test the solution for the second dissipation law given by $\nu_2$ produces a smaller mass--loss rate.  At that choice of critical point, the velocity $v_c$ is larger than that predicted for $\nu_1$ according to Equation \ref{e15}.  Similarly the derivative of the velocity at the critical point is  larger than in the case of the first dissipation law.  At the star's surface, $r = R_*$, the dissipation law does not play a role, implying the momentum equation and the quantity $v(R_*)v^\prime(R_*)$ is approximately the same for both cases. Hence there are three possibilities for the velocity at the surface: the velocity for the second dissipation law is less than, equal, or greater than the velocity for the first dissipation law. If the velocity given by $\nu_2$ is less, then the velocity derivative at the surface is greater, which causes the velocity as a function of distance from the stellar surface to increase too rapidly and not satisfy the Singularity and Regularity conditions at $r_c$.  If the velocity at the surface for $\nu_2$ is greater than or equal to that in the first case then the density structure given by $\rho \propto \dot{M}/v^2$ is smaller at all distances. The continuum optical depth is less than $2/3$ for the second dissipation law and thus the critical point $r_c$ that would satisfy the second dissipation is at a larger distance from the star than the critical point in the first case.   The larger value of $r_c$ for the case of $\nu_2$ increases the mass--loss rate and acts to cancel the effect of greater dissipation. 

The conclusion of these two sections is the dissipation power law will have a significant effect on density and velocity structure but only a minor effect on the mass--loss rate.  The effect on mass loss due to  dissipation given by the power law is cancelled by forcing a different value of the critical point to satisfy the requirement that the continuum optical depth of the wind is $2/3$.  So far, the dissipation exponent $\nu$ has been treated as a variable where specific cases have been explored, $\nu = 0,2,3$.  The results show that $\nu$, while not greatly affecting the mass--loss rate, causes complications regarding the Parker point, $r_p$, except for the case of $\nu = 2$.  Therefore, to avoid those pitfalls, the mass loss of pulsating winds will be treated with the dissipation law given by $\nu = 2$. 

\subsection{Comparison of Pulsating and Non--Pulsating Winds}
It is useful to compare the instantaneous mass--loss rate at some phase of pulsation to the mass--loss rate predicted by only radiative driving in a simple case.  In this case, one can consider the region of the wind where $v >> a$, ignoring terms of the order $a^2/v^2$  and assuming  the term $2a^2r << GM_*(1-\Gamma_e) - \zeta R_*^2$ (again $\nu$ is assumed to equal 2). These assumptions still allow the solution to probe most of the subcritical region $r<r_c$ as well as the supercritical region $r>r_c$ to large $r$ \citep{Lamers1999}, but does limit the solution from describing the case where the $ GM_*(1-\Gamma_e) - \zeta R_*^2 \approx 0$. Thus it is possible to rewrite Equation \ref{e13} as
\begin{equation}\label{e41}
r_cv_cv_c^\prime \approx \left(\frac{\alpha}{1 - \alpha}\right)\left[GM_*(1-\Gamma_e) - \zeta R_*^2\right].
\end{equation}
If it is assumed that the right hand term, over the range specified, does not vary then $r_cv_cv_c^\prime$ is approximately constant meaning $rvv^\prime$ is constant over that range.  Therefore the velocity structure of the wind ejected at some instant can be approximated as
\begin{equation}\label{e42}
 v^2 = v_c^2 + 2\left(\frac{\alpha}{1-\alpha}\right)\left[GM_*(1-\Gamma_e) - \zeta R_*^2\right] \left(\frac{1}{r_c} - \frac{1}{r}\right).
\end{equation}
Furthermore the term $v_c^2$ can be replaced with the approximate form of Equation \ref{e15}, such that the velocity as a function of $r$ is
\begin{equation}\label{e43}
v^2 = \left(\frac{\alpha}{1-\alpha}\right) \left\{ \left[\frac{GM_*(1-\Gamma_e)}{r_c} - \frac{\zeta R_*^2}{r_c}\right]  + 2\left[GM_*(1-\Gamma_e) - \zeta R_*^2\right]\left(\frac{1}{r_c} - \frac{1}{r}\right) \right\}.
\end{equation}
Rearranging and solving for $r_c$ at the surface $r = R_*$ and taking $v(R_*) \approx a$
\begin{equation}\label{e44}
r_c = \frac{3}{2} R_*\left[1 + \frac{1-\alpha}{\alpha}\frac{a^2}{v^2_{\rm{esc, eff}}}\right]^{-1}
\end{equation}
where $v^2_{\rm{esc, eff}} = GM_*(1 - \Gamma_e)/R_* - \zeta R_*$.   This result is similar to the solution of the critical radius for radiative driving from \cite{Lamers1999}, except the escape velocity is different.  

When the function $\zeta$ is greater than zero then the effective escape velocity for the pulsating case is smaller than the escape velocity when there is no pulsation  implying $r_c$ in the case of pulsating stars is smaller than the critical point for no pulsation.  Therefore in the quasi--static pulsating limit one would expect the wind to have a larger acceleration than in the case of only radiative driving.  The instantaneous mass--loss rate in the approximate case, using the assumptions given above with Equation \ref{e17}, is 
\begin{equation}\label{e45}
\dot{M} = \left(\frac{\sigma_e L_* k }{4\pi c}\frac{Z}{Z_\odot}(1-\alpha)\right)^{1/\alpha}\left(\frac{4\pi}{\sigma_e v_{th}}\right)  \left(\frac{\alpha}{1-\alpha}\right)\left[GM_*(1-\Gamma_e)  - \zeta R_*^2\right]^{1-1/\alpha} .
\end{equation}
Relative to the approximate mass--loss rate due to radiative driving only, the ratio of mass--loss rates is
\begin{equation}\label{e46}
 \frac{\dot{M}_{\rm{puls}}}{\dot{M}_{\rm{rad}}} \approx \left[\frac{\sigma_e(t)}{\bar{\sigma}_e}\right]^{-1 + 1/\alpha}\left[\frac{L_*(t)}{\bar{L}_*}\right]^{1/\alpha}\left(\frac{v_{\rm{th}}}{\bar{v}_{\rm{th}}}\right)^{-1} \left[1- \frac{\zeta R_*^2}{GM_*(1-\Gamma_e)}\right]^{1-1/\alpha},
\end{equation}
where the quantities denoted with a bar are time averaged over one period of pulsation.  It is interesting to note the instantaneous mass--loss rate scales directly with the variations of luminosity to the power of $1/\alpha$, and the electron scattering opacity to the power $-1 + 1/\alpha$ and inversely with the square root of the temperature through the thermal velocity. More important is the last term in square brackets containing the ratio of the acceleration due to shocks and pulsation and the effective gravitational acceleration.  The exponent $1 - 1/\alpha$ is less than zero as $\alpha$ is of order $1/2$, meaning as $\zeta$ becomes larger so will the instantaneous mass--loss rate.  Furthermore it is clear when $\zeta$ is of order the effective gravity the mass--loss rate goes to infinity; however, this regime  violates the assumption $2a^2r << GM_*(1-\Gamma_e) - \zeta R_*^2$.  Still this does illustrate the non--linear behavior of the mass--loss rate due to the added effects of pulsation and shocks.  The goal now is to quantify how large the Cepheid mass--loss rates can be.

\section{Testing the Wind Model with Cepheid Data}
\begin{table*}
\begin{center}
\begin{footnotesize}
\begin{tabular}{lccccccc}

\hline
Name & Period(d)& M$_V$& $\Delta$V & Radius (R$_\odot$)& $\Delta$R/R & Mass (M$_\odot$) & T$_{\rm{eff}}$(K)\\
\hline
R TrA & 3.388 & -2.74 & 0.561 & 25.3 & 0.074 & 2.4 & 6470\\
RT Aur & 3.724 & -2.86 & 0.803 & 35.1 & 0.103 & 4.7 & 5640\\
BF Oph & 4.064 & -2.97 & 0.636 & 34.5 & 0.101 & 3.9 & 5840\\
V Vel & 4.375 & -3.06 & 0.689 & 32.8 & 0.116 & 3.1 & 6110 \\
T Vul & 4.436 & -3.08 & 0.643 & 38.2 & 0.100& 4.5 & 5690\\
V482 Sco &  4.529 & -3.11 & 0.652 & 44.4 & 0.103 & 6.4 & 5320\\
S Cru & 4.688 & -3.15 & 0.690 & 42.1 & 0.108 & 5.3 & 5510\\
AP Sgr & 5.058 & -3.25 & 0.832 & 44.0 & 0.110 & 5.3 & 5510\\
V350 Sgr & 5.152 & -3.27 & 0.705 & 47.8 & 0.110 & 6.3 & 5320\\
$\delta$ Cep & 5.370 &-3.32 &  0.838 & 41.6 & 0.116 & 4.2 & 5800\\
V Cen  & 5.495  & -3.35 & 0.804 & 45.3 & 0.119 & 5.0 & 5820\\
Y Sgr & 5.768 & -3.41 & 0.725 & 50.0 & 0.120 & 6.0 & 5370\\
RV Sco &  6.067 & -3.47 & 0.824 & 47.7 & 0.102 & 4.9 & 5570\\
S TrA & 6.324 & -3.53 &  0.768 & 39.2 & 0.106 & 2.8 & 6230\\
AW Per &  6.456  & -3.55 & 0.812 & 47.3  & 0.118 & 4.4 & 5700\\
BB Sgr  & 6.637 & -3.59 & 0.597 & 40.3 & 0.097 & 2.8 &6230 \\
AT Pup & 6.668 & -3.59 & 0.904 & 45.4 & 0.120 & 3.8 & 5870\\
V Car & 6.699 & -3.60 & 0.601 & 40.6 & 0.094 & 2.8 & 6220\\
U Sgr & 6.745 & -3.61 & 0.717 & 51.4 & 0.115 & 5.1 & 5540\\
V496 Aql & 6.808 & -3.62 & 0.349 & 45.5 & 0.067 & 3.7 & 5900\\
X Sgr  & 7.014  & -3.66 & 0.590 & 49.8 & 0.093 & 4.4 & 5720\\
U Aql & 7.031 & -3.66 & 0.757  & 54.7 & 0.116 & 5.6 & 5440\\
$\eta$ Aql & 7.178 & -3.69 & 0.799 & 54.9  & 0.121 & 5.5 & 5510\\
W Sgr & 7.603 & -3.76 & 0.805 & 63.3 & 0.121 & 7.2 & 5200\\
RX Cam & 7.907  & -3.81 & 0.729 & 76.0 & 0.120 & 10.8 & 4770\\
W Gem & 7.907 & -3.81 & 0.822 & 50.7 & 0.126 & 3.9 & 5840\\
U Vul & 7.998 & -3.82 & 0.718 & 56.5 &  0.098 & 5.0 & 5550\\
GH Lup & 9.268 & -4.01 & 0.192 & 58.2 & 0.040 & 4.3 & 5710\\
S Mus & 9.660 & -4.06 & 0.500 & 71.3 & 0.109 & 6.8 & 5220\\
S Nor & 9.750 & -4.07 & 0.640 & 66.4 & 0.119 & 5.6 & 5420 \\
$\beta$ Dor & 9.840 & -4.08 & 0.630 & 64.4 & 0.118 & 5.1 & 5550\\
$\zeta$ Gem & 10.139 & -4.12 & 0.480 & 64.9 & 0.099 & 5.0 & 5590\\
XX Cen & 10.965 & -4.22 & 0.924 & 57.8 & 0.140 & 3.3 & 6020\\
RX Aur & 11.614 & -4.29 & 0.664 & 63.4 & 0.122 & 3.8 & 5840\\
TT Aql & 13.740 & -4.51 & 1.082 & 95.8 & 0.196 & 8.5 & 5020\\
X Cyg & 16.368 & -4.73 & 0.986 & 118.1& 0.216 & 11.1 & 4760  \\
Y Oph & 17.139 & -4.78 & 0.483 & 93.5 & 0.080 & 5.8 & 5410\\
VY Car & 19.011& -4.91 & 1.065 & 108.9 & 0.220 & 7.3 & 5140\\
RZ Vel & 20.417 & -5.00 & 1.181 & 111.7 & 0.243 & 7.0 & 5210\\
T Mon & 27.040 & -5.36 & 1.028 & 130.6 & 0.230 & 6.8 & 5230\\
l Car  & 35.563 & -5.70 & 0.725 & 183.7 & 0.191 & 10.7 & 4780 \\
U Car & 38.726  & - 5.81 & 1.165 & 162.2  & 0.238 & 6.9 & 5200\\
RS Pup  & 41.400 & -5.90 & 1.105 & 197.7 & 0.246 & 10.3 & 4820\\
SV Vul & 44.978 & -6.00 & 1.054 & 235.5 & 0.246 & 14.2 & 4520\\
\hline
\end{tabular}
\end{footnotesize}
\end{center}
\caption[]{Data for modeling the mass--loss rate behavior of Cepheids. See text for description and references.}
 \label{tab2}
\end{table*}
\begin{table*}[t]
\begin{center}
\begin{tabular}{lcclcc}
\hline
Name &$ \dot{M}_{\rm{puls}} (M_\odot /yr)$&  $ \dot{M}_{\rm{rad}}(M_\odot /yr)$ & Name &$ \dot{M}_{\rm{puls}} (M_\odot /yr)$& $ \dot{M}_{\rm{rad}}(M_\odot /yr)$\\
\hline
R TrA  & $ 4.9 \times 10^{-10}$ &$1.2\times 10^{-11}$&$\eta$ Aql &$1.8 \times 10^{-10}$&$3.6\times 10^{-11}$\\
RT Aur  & $2.3\times 10^{-8}$ &$7.5\times 10^{-12}$&W Sgr &$1.3\times 10^{-10}$&$3.0\times 10^{-11}$\\
BF Oph &$5.4\times 10^{-10}$ &$1.1\times 10^{-11}$&RX Cam & $8.4\times 10^{-11}$&$2.0\times 10^{-11}$\\
V Vel & $1.3 \times 10^{-8}$&$1.7\times 10^{-11}$&W Gem & $3.0 \times 10^{-10}$& $6.2\times 10^{-11}$ \\
T Vul  & $1.3\times 10^{-10}$&$1.2\times 10^{-11}$&U Vul &$ 1.3\times 10^{-10}$&$4.8\times 10^{-11}$ \\ 
V482 Sco & $1.0\times 10^{-10}$ &$9.0 \times 10^{-12}$&GH Lup  & $1.4 \times 10^{-10}$ &$8.3\times 10^{-11}$\\
S Cru & $ 2.1 \times 10^{-10}$ &$1.2\times 10^{-11}$&S Mus &$ 1.4\times 10^{-10}$&$ 5.6\times 10^{-11}$ \\
AP Sgr & $1.5\times 10^{-10}$ &$1.4\times 10^{-11}$& S Nor &$2.0\times 10^{-10}$& $7.0\times10^{-11}$ \\
V350 Sgr &$1.2\times 10^{-10}$& $1.2\times 10^{-11}$& $\beta$ Dor &$2.4 \times 10^{-10}$&$8.3\times 10^{-11}$\\
$\delta$ Cep & $3.6\times 10^{-10}$&$2.3\times 10^{-11}$&$\zeta$ Gem &$2.2 \times 10^{-10}$&$9.5\times 10^{-11}$\\
V Cen & $1.7\times 10^{-9}$&$2.7\times 10^{-11}$& XX Cen & $7.3\times 10^{-10}$& $1.7\times 10^{-10}$\\
Y Sgr &$1.9\times 10^{10}$&$1.7\times 10^{-11}$&RX Aur & $4.6\times 10^{-10}$& $1.7\times 10^{-10}$ \\
RV Sco & $9.3 \times 10^{-11}$& $2.4\times 10^{-11}$&TT Aql &$8.0\times10^{-9}$&$1.1\times 10^{-10}$ \\
S TrA & $2.3 \times 10^{-10}$ &$5.0\times 10^{-11}$&X Cyg &$7.7\times 10^{-8}$&$1.3\times10^{-10}$\\

AW Per & $1.8\times 10^{-10} $ &$3.2 \times 10^{-11}$&Y Oph &$5.1\times 10^{-10}$ &$3.0\times 10^{-10}$ \\
BB Sgr & $2.1 \times 10^{-10}$ & $5.7\times 10^{-11}$&VY Car & $4.7\times 10^{-9}$&$2.8\times 10^{-10}$\\
AT Pup & $2.4 \times 10^{-10}$ &$4.1\times 10^{-11}$&RZ Vel &$3.7\times 10^{-8}$&$3.8\times 10^{-10}$\\
V Car& $ 2.0 \times 10^{-10}$ &$5.8\times 10^{-11}$&T Mon &$5.5\times 10^{-9}$&$8.1\times 10^{10}$\\
V496 Aql & $ 1.0 \times 10^{-10}$ &$4.5\times 10^{-11}$ &$l$ Car & $2.4\times 10^{-9}$&$9.9\times 10^{-10}$ \\
X Sgr &$1.2\times 10^{-10}$&$4.1\times 10^{-11}$ &U Car &$1.0 \times 10^{-8} $&$2.0 \times 10^{-9}$\\
U Sgr &$1.4\times 10^{-10}$&$3.1\times 10^{-11}$&RS Pup &$ 6.5\times 10^{-9}$&$1.6\times 10^{-9}$\\
U Aql &$ 1.3\times 10^{-10}$&$3.1\times 10^{-11}$ &SV Vul & $4.8\times 10^{-9}$ &$1.3\times 10^{-9}$ \\
\hline
\end{tabular}
\end{center}
\caption{Predicted mass--loss rates for Cepheids using shock and pulsation dynamics along radiative driving and assuming only radiative driving.} 
\label{t3}
\end{table*}

The Cepheid instability strip spans a large range of mass, radius and luminosity, so to understand the role of mass loss, it is necessary to have a significant number of Cepheid models and observations.  That information can be used to calculate mass--loss rates across the instability strip and probe how mass loss evolves with time.  The circumstellar shells that have been observed can be modeled as material from the wind that condenses to dust when the temperature reaches approximately $1500$ K, as mentioned by \cite{Kervella2006}.  The dust would absorb light and re-emit isotropically, some of the radiation is toward the observer and thereby causing a flux excess at near infrared and longer wavelengths.

To calculate mass--loss rates using the modified version of the CAK method, it is necessary to know the following parameters: the radius, amplitude of radius variation, luminosity, amplitude of luminosity variation, mass, and the period, along with the metallicity.  Furthermore the values of $\alpha$ and $k$ are needed for radiative driving.  These parameters are based on the effective temperature and gravity of the star.  For this work the parameters chosen are: $\alpha = 0.465$ and $k = 0.064$ from the analysis of \cite{Abbott1982}.  These values are used for models with temperatures ranging from $4600$ K to $6000$ K.  There is a lack of parameters calculated at effective temperatures less then $6000$ K, but $\alpha$ and $k$ are not likely to vary much since the main elements that drive the mass loss radiatively are iron, neon and calcium as well as  contributions from hydrogen and helium remain the same \citep{Lamers1999}. Molecules may play a small role as molecular opacities are important in AGB star winds, $T_{\rm{eff}} \approx 3000$ K \citep{Helling2000}, but they are ignored in this analysis.

The amplitude of the radius variation can be determined by integrating the radial velocity profile over one period.  \cite{Moskalik2005} compiled a list of viable interferometric targets and calculated the amplitudes of radius variation. The Cepheids on this list that pulsate in the fundamental mode are used here.  Since all of the Cepheids in the sample are galactic stars the metallicity is assumed to be solar,$Z=0.02$, consistent with the shock model from \cite{Fokin1996}.  Recently, \cite{Asplund2005} have suggested that the solar metallicity might be $Z_\odot = 0.012$, This downward revision is being questioned because it is incompatible with helioseismology.  It is unclear how the lower metallicity would affect the shock structure in the atmosphere of a Cepheid, although if one increases the hydrogen and helium abundances then the shocks generated in the hydrogen and helium ionization fronts might have more energy.  An obvious consequence of a lower metallicity is seen in Equation \ref{e45}, where the mass--loss rate explicitly depends on the metallicity, and would lead to the mass--loss rate at lower $Z$ being $1/3$ the mass--loss rate at $Z = 0.02$. However, it is shown other parameters affect the mass loss more significantly.  The radius, luminosity and luminosity variation are found in the David Dunlap Observatory Catalogue of Galactic Classical Cepheids \citep{Fernie1995}.  The effective temperature iscalculated from the mean luminosity and radius while the mass is calculated using the Period--Mass--Radius relation from \cite{Gieren1989} and \cite{Fricke1972}:
\begin{equation}\label{e47}
M_* = [40P(\mbox{days})]^{-1.49}(R_*/R_\odot)^{2.53}M_\odot.
\end{equation}
The properties of the observed Cepheids are listed in Table \ref{tab2} and the computed mass--loss rates are listed in Table \ref{t3}.

\begin{figure*}[t]
	\begin{center}
		\epsscale{0.48}
		\plotone{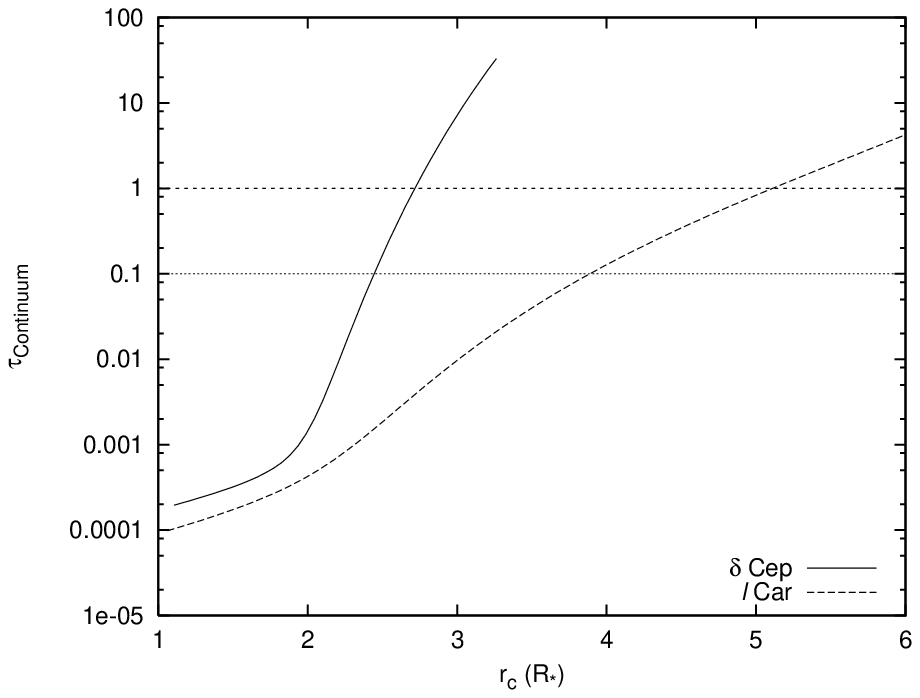}
		\plotone{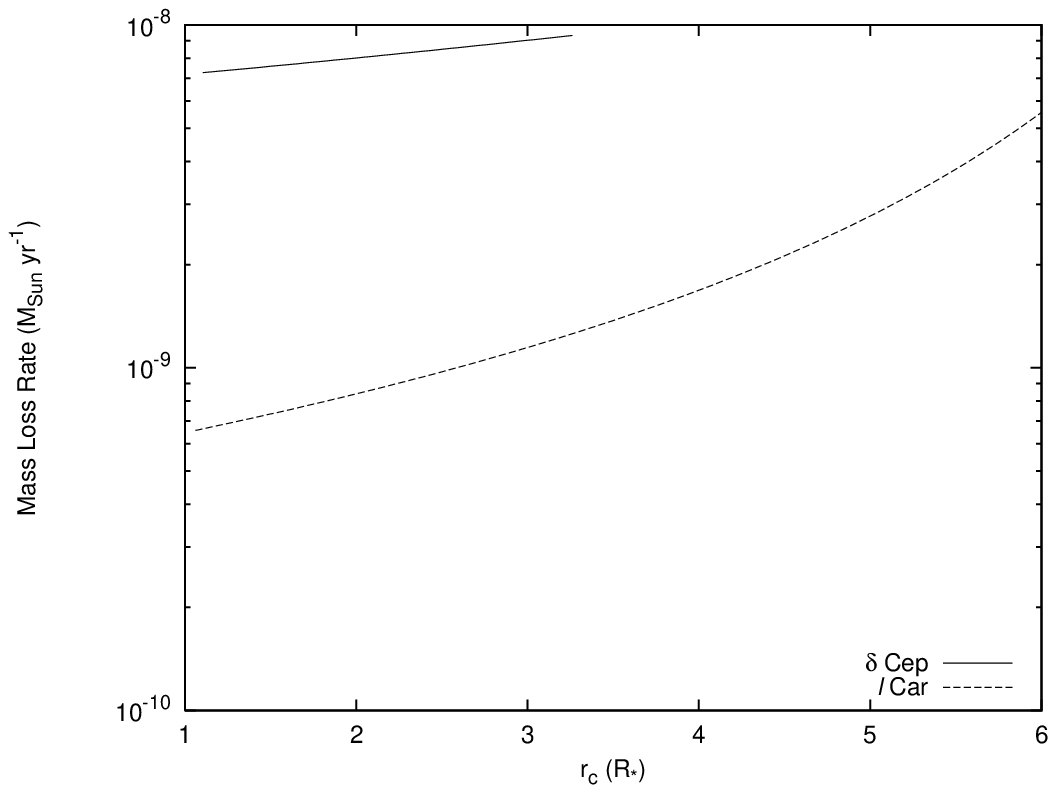}
	
		\caption{(Left Panel) The predicted continuum optical depth of the wind for $\delta$ Cep and $l$ Car at minimum radius. The horizontal lines highlight the regime where the $\tau$ is of order $2/3$. (Right Panel) The predicted mass--loss rate for $\delta$ Cep and $l$ Car at minimum radius. The rates do not vary significantly with $r_c$.  }
		\label{f0a}
	\end{center}
\end{figure*}

The first test is to compare the sensitivity of the pulsation mass loss to the value of the critical point, $r_c$.  If the mass loss varies significantly as a function of $r_c$ for a given set of parameters then the predicted mass--loss rate will strongly depend on the criterion that the continuum optical depth of the wind be $2/3$.  The  continuum optical depth is shown in Figure \ref{f0a} (Left) as a function of the critical point, $r_c$, for two Cepheids in the sample, $l$ Car and $\delta$ Cep at minimum radius.  The continuum optical depth varies between $0.1$ and $1$ over a small range of critical points, for $\delta$ Cep the range is approximately $0.5R_*$ and for $l$ Car it is about $1.5R_*$.  Over the two ranges of critical points, the mass--loss rates vary only about a factor of $2$ and $3$ as shown in Figure \ref{f0a} (Right).  Therefore, the predicted mass--loss rate of a Cepheid is not sensitive to the criterion requiring the continuum optical depth of the wind to be $2/3$.

\begin {figure}[t]
      \begin{center}
      	\epsscale{0.75}
              \plotone{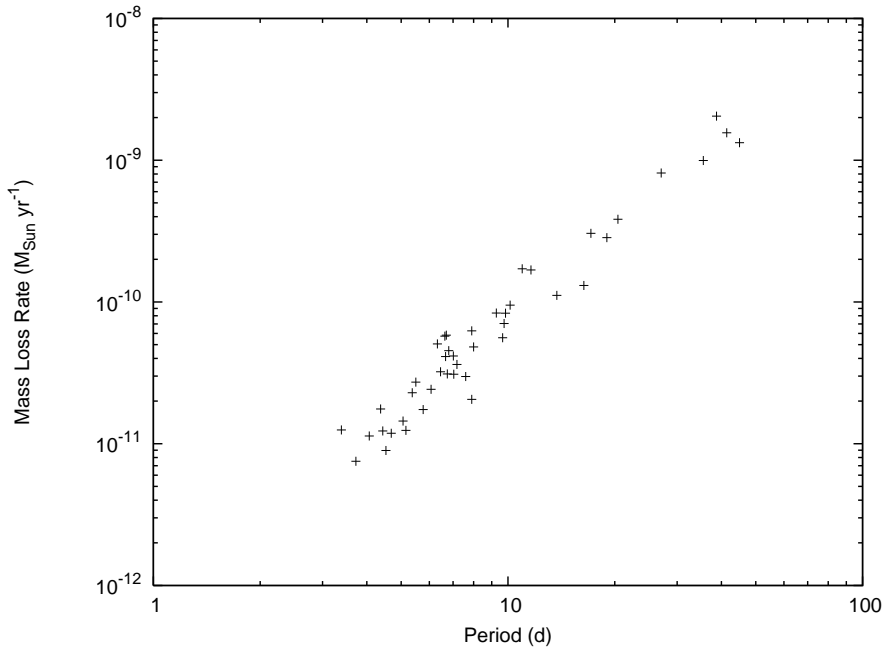}
 \caption{The predicted mass--loss rates for Cepheids in the quasi--static approximation using radiative driving as the only mechanism for generating the wind.  }
        \label{f1}
     \end{center}
\end {figure}  
As a reference, mass--loss rates for this set of observational data are predicted assuming a static stellar surface and the only driving mechanism is radiative with pulsation terms ignored.  In Figure \ref{f1}, the mass--loss rates are plotted as a function of pulsation period.  The linear relation in the Log--Log plot is striking, if one considers the equation for mass loss in the CAK method in the approximate case given in Section 2 and ignoring pulsation, then 
\begin{equation}\label{e48}
\log \dot{M} \approx \left(1 - \frac{1}{\alpha}\right)\log M_*  + \frac{1}{\alpha}\log L  - \frac{1}{2}\log T_{\rm{eff}}+ \rm{Const}.
\end{equation}  
The mass can be written in terms of the period and radius via Equation \ref{e47} and the radius is a function of the effective temperature and luminosity, $R \propto L^{1/2}T^{-2}$.  Furthermore the effective temperature can be represented by the color $(B-V)_0$ using the transform from \cite{Fry1999}.  The radiative driven mass loss is thus
\begin{equation}\label{e48c}
\log \dot{M}_{\rm{rad}} (M_\odot/ yr) = -14.047 + 0.695\log L(L_\odot) - 0.574(B-V)_0 + 1.71\log P(d).
\end{equation}
The luminosity can be eliminated using a Period--Luminosity--Color relation from \cite{DiBenedetto1995} to obtain an $\dot{M}PC$ relation for the radiative driven mass--loss rates for specific Cepheids,
\begin{equation}\label{e48d}
\log \dot{M}_{\rm{rad}} = -12.11 + 2.69\log P - 1.16(B-V)_0.
\end{equation}
This relation is for specific Cepheids but to understand the apparent relation of radiative mass loss as a function of period in Figure \ref{f1}, we wish to eliminate the color dependence to derive a statistical representation of the mass--loss rate.  Returning to Equation \ref{e48c}, the luminosity and the color can be eliminated by using Period--Luminosity and Period--Color relations from \cite{Tammann2003} yielding
\begin{equation}\label{e48e}
\log \dot{M}_{\rm{rad}} = -12.45 + 2.27\log P.
\end{equation}
This relation is a theoretical derivation of the radiative driven mass--loss rates for the statistical sample of galactic Cepheids based on best-fit Period--Luminosity and Period--Color relations.

 Performing a least squares fit on the predicted radiative driven mass--loss rates as a function of the period as shown in Figure \ref{f1}, one finds
\begin{equation}\label{e49}
\log \dot{M}_{\rm{rad}} (M_\odot/yr)= 2.1\log P(d) - 12.2  .
\end{equation}
The two relations differ but this is to be expected. The derived result ignores the contributions due to continuum radiative driving and assumes that $GM_* >> a^2r$.  The best fit relation will have statistical uncertainties as the majority of the Cepheids examined have periods $P < 10d$. Therefore the best--fit and derived relations for the radiative driven mass--loss rate as a function of period agree and provide a rough lower limit for the mass--loss rates of Cepheids. Furthermore this best--fit relation is equivalent to a Reimer's relation $\dot{M} = \eta L_*R_*/M_*$ \citep{Reimers1977}, where all quantities are in solar units and time is in years.  The standard value is $\eta = 10^{-13}$; for Cepheids this would predict mass--loss rates much larger than found here using the CAK method.  Fitting this relation to the radiative driving mass--loss rates determined here, one finds a value of $\eta \approx 4.4\times 10^{-15}$.  
\begin {figure*}[t]
      \begin{center}
\epsscale{0.48}
   \plotone{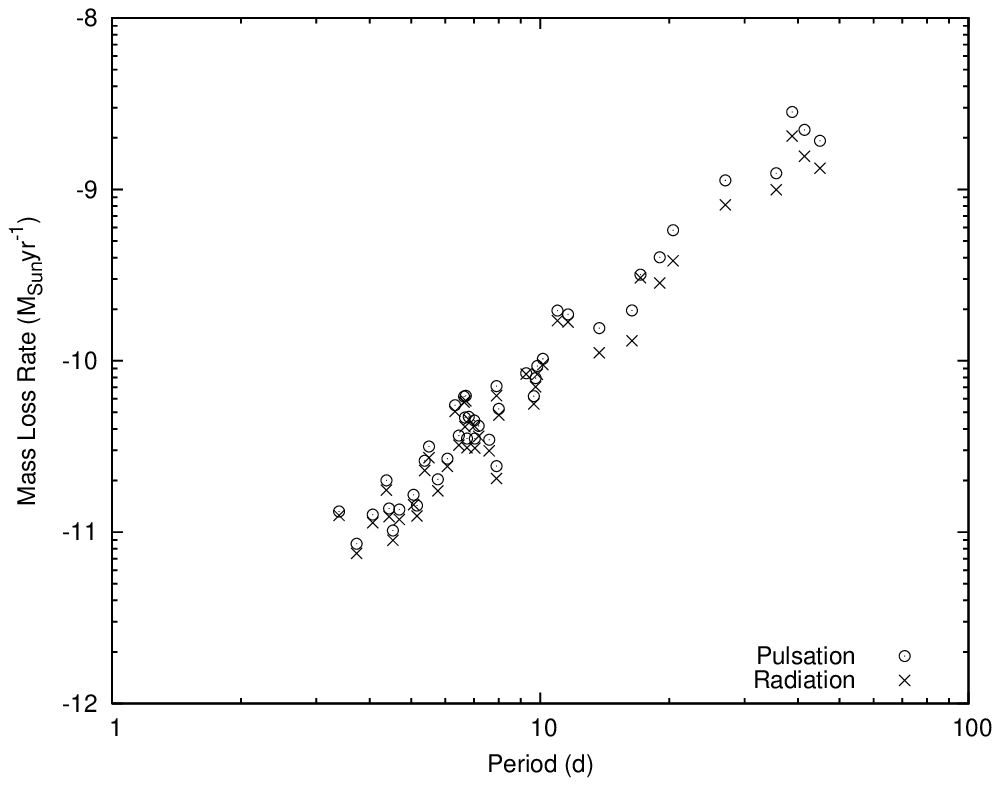}
   \plotone{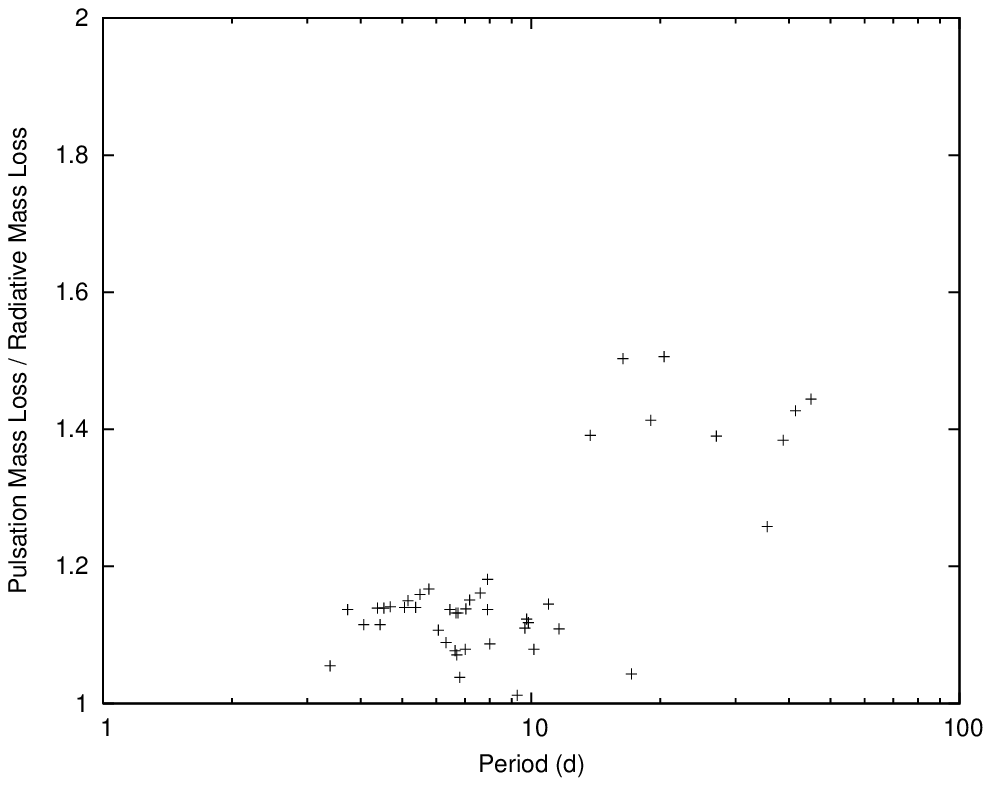}
 \caption{(Left Panel) The mass--loss rates for Cepheids using the combination of radiative driving and pulsation but ignoring shocks to generate the wind plotted with the mass--loss rates from only radiative driving for comparison. (Right Panel) The ratio of the mass--loss rates computed using pulsation and that using only radiative driving.  Accelerating the wind via pulsation does not effect the mass loss significantly by $\lesssim 50\%$. The mass loss is more enhanced for Cepheids with period $> 10$ days is due to the combination of larger amplitudes of radius variation on average and lower gravity.}
        \label{f1a}
     \end{center}
\end {figure*}  

The second step is to compute the mass--loss rates of Cepheids assuming the contribution from shocks is zero and the wind is accelerated by momentum from pulsation and radiation. The result is shown in Figure \ref{f1a} (Left) where the mass--loss rates due to pulsation but not shocks are plotted alongside the mass--loss rates from radiative driving only.  The comparison shows the pulsation does not greatly enhance the mass loss and the result is further highlighted in Figure \ref{f1a} (Right) where the ratio of the two mass--loss rates is also shown.  Pulsation does amplify the mass--loss rates but it is a $50\%$ effect at most, which will not produce significant mass loss.  The result is not surprising if one considers Equation \ref{e46} and replaces $\zeta$ with just $\omega^2\Delta R$. Cepheid pulsation can be approximated as a linear perturbation implying the acceleration due to pulsation is much smaller than the gravity of the Cepheid. In that case, the ratio of pulsation mass--loss rates to the radiatively driven mass--loss rates is close to unity. Therefore it is necessary to consider the effect of shocks on the wind.

\begin {figure*}[t]
      \begin{center}
     \epsscale{0.48}
	\plotone{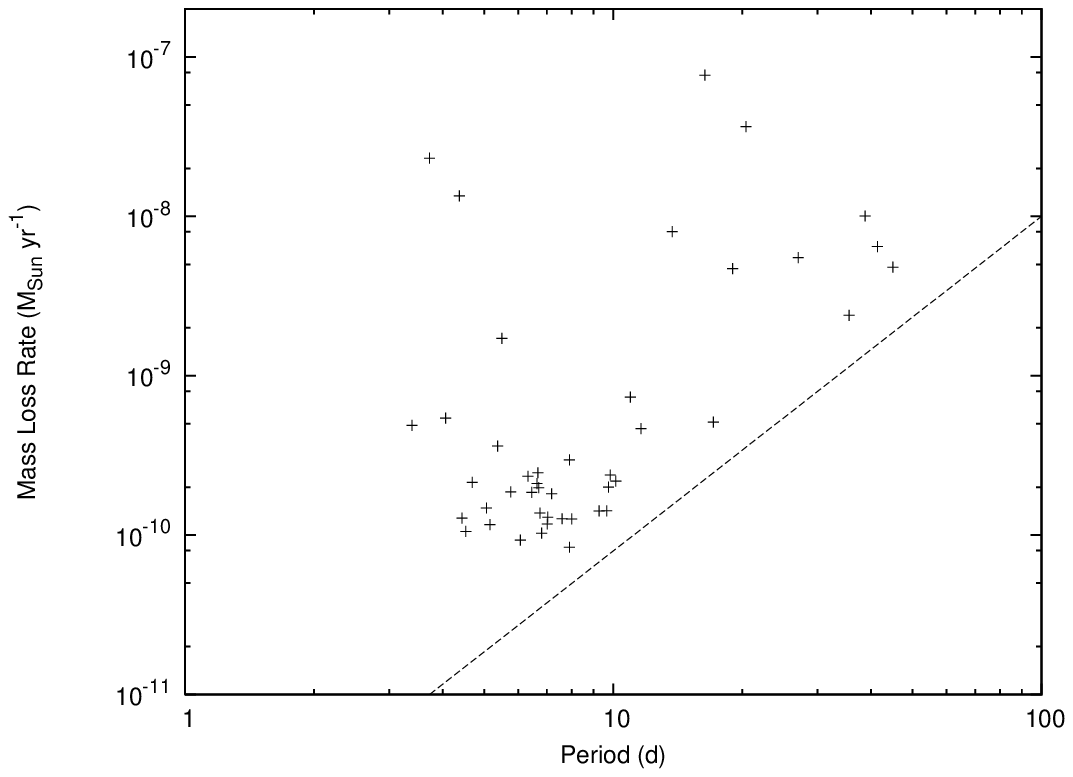}
	\plotone{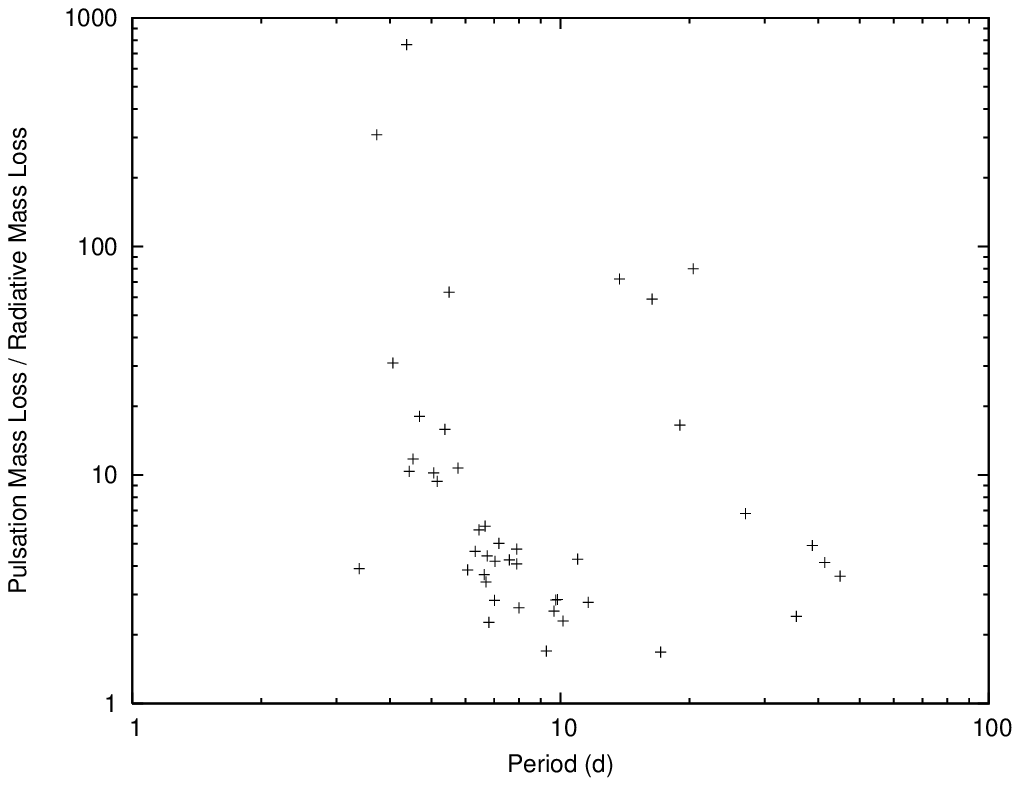}
 \caption{(Left Panel) The mass--loss rates for Cepheids using the combination of radiative driving, pulsation and shocks to generate the wind. The dotted line is the least square fit to the radiative driven mass--loss rates shown in Figure \ref{f1}. (Right Panel) A comparison of the mass--loss rates computed using pulsation plus shock effects and that using only radiative driving.  The wind is strongly enhanced for some Cepheids with the largest enhancement being a factor of approximately $750$ and the smallest about $1.7$ times.  }
        \label{f2}
     \end{center}
\end {figure*}

When the terms describing the shocks and pulsation are included in the calculation, the mass--loss rates change dramatically.  This can be seen in Figure \ref{f2} (Left) where the pulsation mass loss is plotted as a function of period.  The mass--loss rates do not follow a simple relation such as a $\dot{M}-P$ relation for radiatively driven mass loss or a Reimer's relation.  Some Cepheids have very large mass--loss rates compared to other Cepheids, which can be understood from the approximate version of the mass--loss rate, Equation \ref{e47}. The enhanced mass--loss rate, shown in Figure \ref{f2} (Right), depends on $(1 - \zeta/g)^{-1}$ if one assumes $\alpha \approx 1/2$, where $g$ is the gravity of the star;  the term is maximized when either $\zeta$ is large or $g$ is small. The gravity is smallest when the Cepheid is at maximum expansion, $\phi = 0.5$, but the acceleration due to pulsation is at a minimum and the shock amplitude is zero.  Therefore $\zeta$ at this phase, is acting to decrease the mass--loss rate. The pulsation plus shock function, $\zeta$, is largest at minimum radius where the shock amplitude is large and the acceleration due to pulsation is maximum, meaning both effects contribute.  The sum of the two accelerations, which act at similar magnitudes, and the fact that the pulsation depends on the amplitude $\Delta R$ implies there is no simple formula for the mass--loss rate.

\begin{figure*}[t]
   \begin{center}
   	\epsscale{0.48}
	\plotone{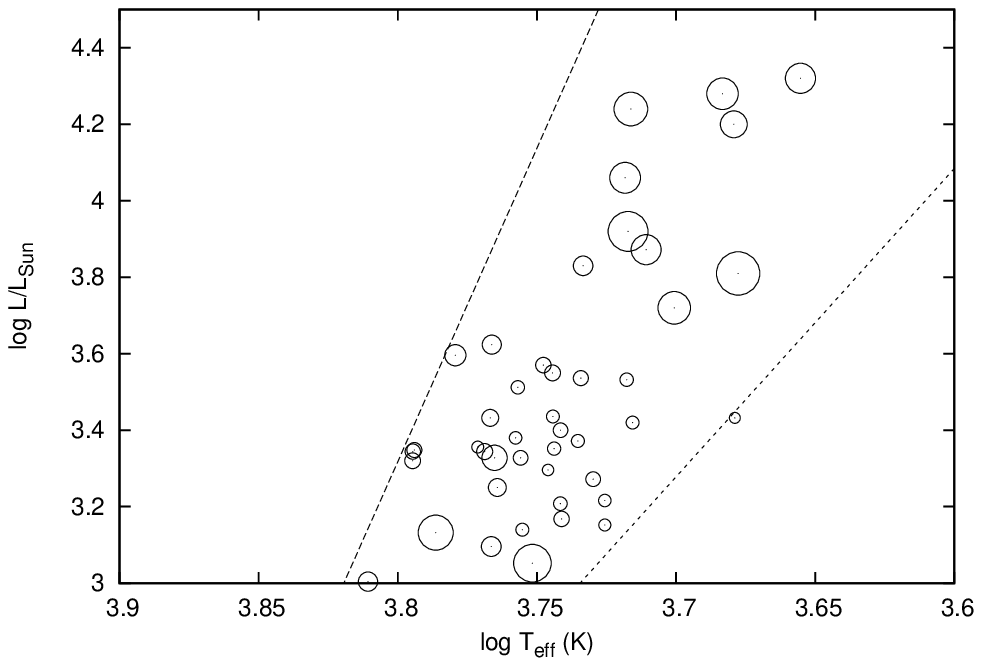}
	\plotone{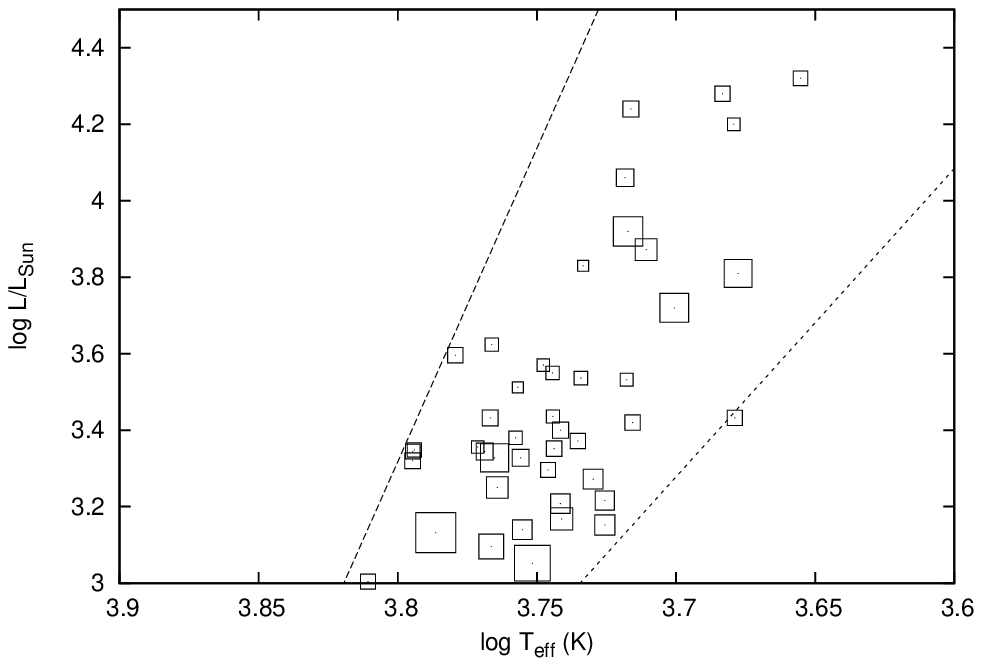}
   \end{center}
 \caption{The observed Cepheids plotted on a Cepheid instability strip of the HR diagram where the dashed lines represent approximate boundaries. (Left Panel) The size of the circles represent the calculated mass--loss rates for the Cepheids and (Right Panel) the size of the squares represent the ratio of the pulsational mass loss and the radiative driven mass loss.  All of the brighter Cepheids, $\log L/L_\odot \ge 3.7$, have large mass--loss rates, while this is true for only a few of the less bright Cepheids  but only a fraction of the luminous Cepheids have their mass loss significantly increased due to pulsation and shocks. }
         \label{f4}
\end{figure*}

Only some Cepheids have a dramatic enhancement of mass loss, implying there are restricted regions in the instability strip where Cepheids are more susceptible to lose mass. Figure \ref{f4} (Left) shows the location of the Cepheids modeled here with the size of the symbols representing the amount of mass loss.
The plot shows that larger Cepheids existing in the upper part of instability strip exhibit higher mass--loss rates, as would be expected from the calculations of radiative driving. It is also interesting that the few short period Cepheids with high mass--loss rates are scattered along the effective temperature axis but have consistently lower luminosity, which could be a result of a lower mass and short period contributing to lower gravity and higher pulsation plus shock acceleration respectively.  Figure \ref{f4} (Right) shows the Cepheids on the HR diagram but with the point sizes now representing the amount of mass loss enhancement.   The results are striking; there appears to be two bands where the mass loss is enhanced. This may imply the mass loss may be related to which crossing of the instability strip the Cepheid is making.

\begin{figure}[t]
      \begin{center}
      \epsscale{0.75}
  	\plotone{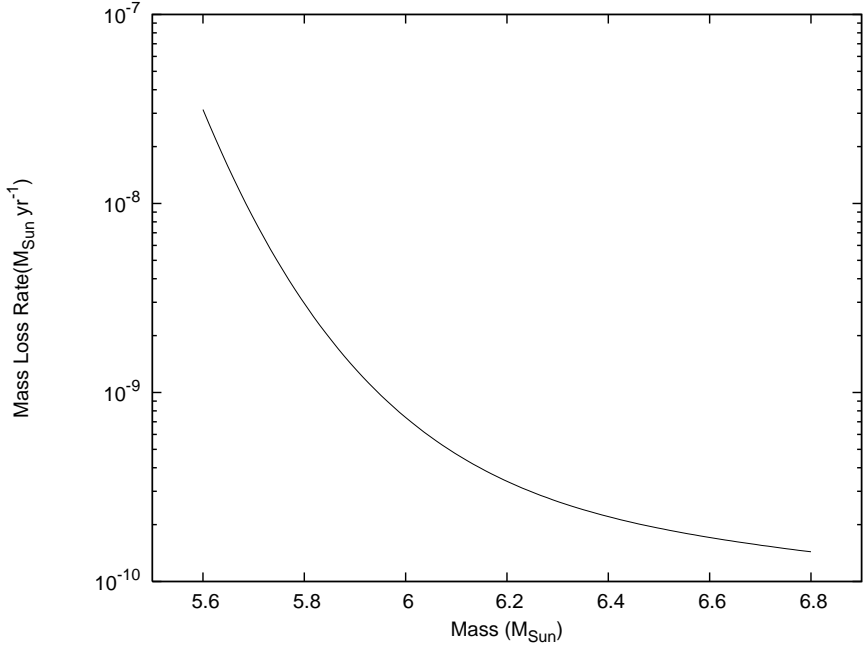}
 \caption{The predicted mass--loss rate for S Mus found by holding all necessary parameters describing the Cepheid constant except for the mass, which is allowed to vary. }
        \label{f12}
     \end{center}
\end {figure}

There are many sources of uncertainty in this analysis. The most important source is the mass for each Cepheid,  based on a Period--Mass--Radius relation, which has scatter.  Deviations of the mass in the relation to the observed values are important.  Consider the Cepheid S Mus, for which a mass of $6.8M_\odot$ is used here; however \cite{Evans2006} find S Mus has a mass of $6.0\pm 0.4M_\odot$.  It is clear from Equation \ref{e46} that a lower mass will increase the rate of mass loss if all other parameters remain the same. One can test the dependence of the mass--loss rate on the mass of the Cepheid by calculating a set of models with varying mass and holding all other parameters the same.  The effect is shown in Figure \ref{f12} for S Mus, ranging the mass from $5.6M_\odot$ to $6.8M_\odot$. For the larger masses the mass--loss rate is not sensitive to the mass, at $6.8M_\odot$, $\dot{M} = 1.4\times 10^{-10}M_\odot/yr$ and increases by only a factor of five at $6.0M_\odot$.  At lower masses, however, the mass--loss rate exhibits a non--linear dependence on the mass increasing by two orders of magnitude from $M = 6.0$ to $M = 5.6M_\odot$.  Therefore when the mass--loss rates are large, it is important to have strong constraints on the mass.

The uncertainty in the calculation may be quantified by considering the analytic derivation of the error of the pulsation--driven mass--loss rate based on Equation \ref{e17}, simplified by assuming the quantities $a^2r_c$ and $\Gamma_e$ are insignificant.  Writing the error as $\Delta F(x_i) = \sqrt{\sum_i (\partial F/\partial x_i)^2\sigma_{x_i}^2}$ where $\sigma_{x_i}$ is the error of each quantity,
\begin{equation}\label{e49a}
\frac{\sigma_{\dot{M}} }{\dot{M}}= \left[ \left(\frac{1}{\alpha L_*}\right)^2 \sigma^2_L + \left(\frac{1}{2T_{eff}}\right)^2\sigma_T^2 + \frac{G^2\sigma^2_M + R_*^4\sigma_\zeta^2 + 4\zeta^2R_*^2\sigma_R^2}{(GM_* - \zeta R_*^2)^{2}}\right]^{1/2}.
\end{equation}
In this equation, it is also assumed that the uncertainty of the period and amplitude of luminosity variation is negligible.  The uncertainty of the function $\zeta$ is given by
\begin{equation}\label{e49b}
\sigma_\zeta = \left\{[ \omega^2 \sigma_{\Delta R}]^2 + \left[\frac{du_\delta}{Pd\phi}\left(\frac{L}{L_\delta}\right)^{1/8}\left(\frac{P}{P_\delta}\right)^{-7/24}\left(\frac{M}{M_\delta}\right)^{5/12}\right]^2\left[\left(\frac{1}{8}\frac{\sigma_L}{L_*}\right)^2 + \left(\frac{5}{12}\frac{\sigma_M}{M_*}\right)^2\right]\right\}^{1/2}.
\end{equation}
Upon considering the third term of Equation \ref{e49a}, the uncertainty of the mass--loss rate is inversely dependent on the balance of forces $GM_* - \zeta R^2_*$.  It was shown in Section 2 that the pulsation driven mass--loss rate is large when the balance of forces is small and likewise $\sigma_{\dot{M}}/\dot{M}$ is also inversely proportional to the balance of forces.  This implies that when the mass--loss rate is large due to pulsation and shocks then the error is even larger.  In this case, the uncertainty may be larger by more than an order of magnitude depending on the uncertainty of the mass, radius, luminosity, and amplitude of radius variation.  The uncertainty in the effective temperature is not a significant source of error.   This implies that the calculation is very sensitive to the values of the parameters used and reiterates the need for more precise values of mass, as well as luminosity and radius.  

\begin{figure}[t]
      \begin{center}
  	\plottwo{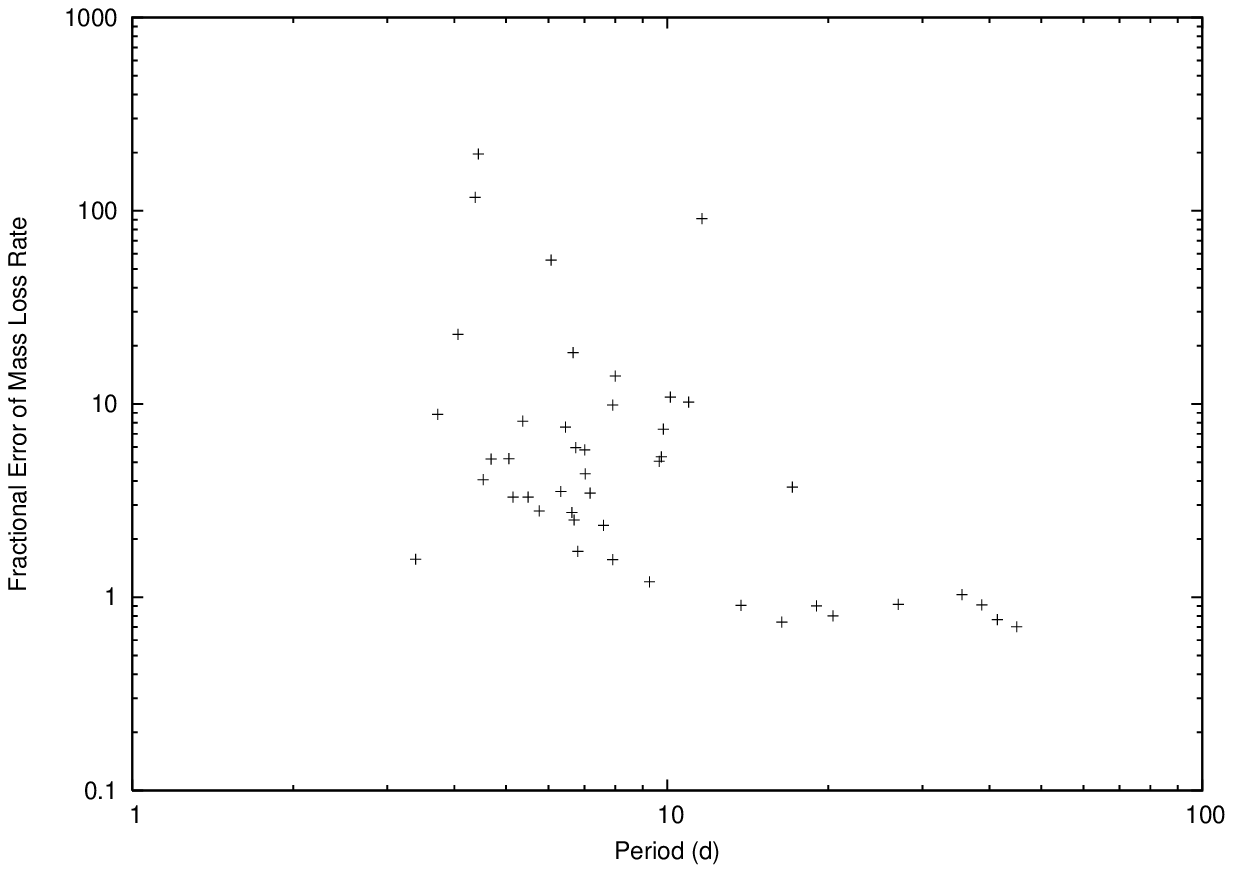}{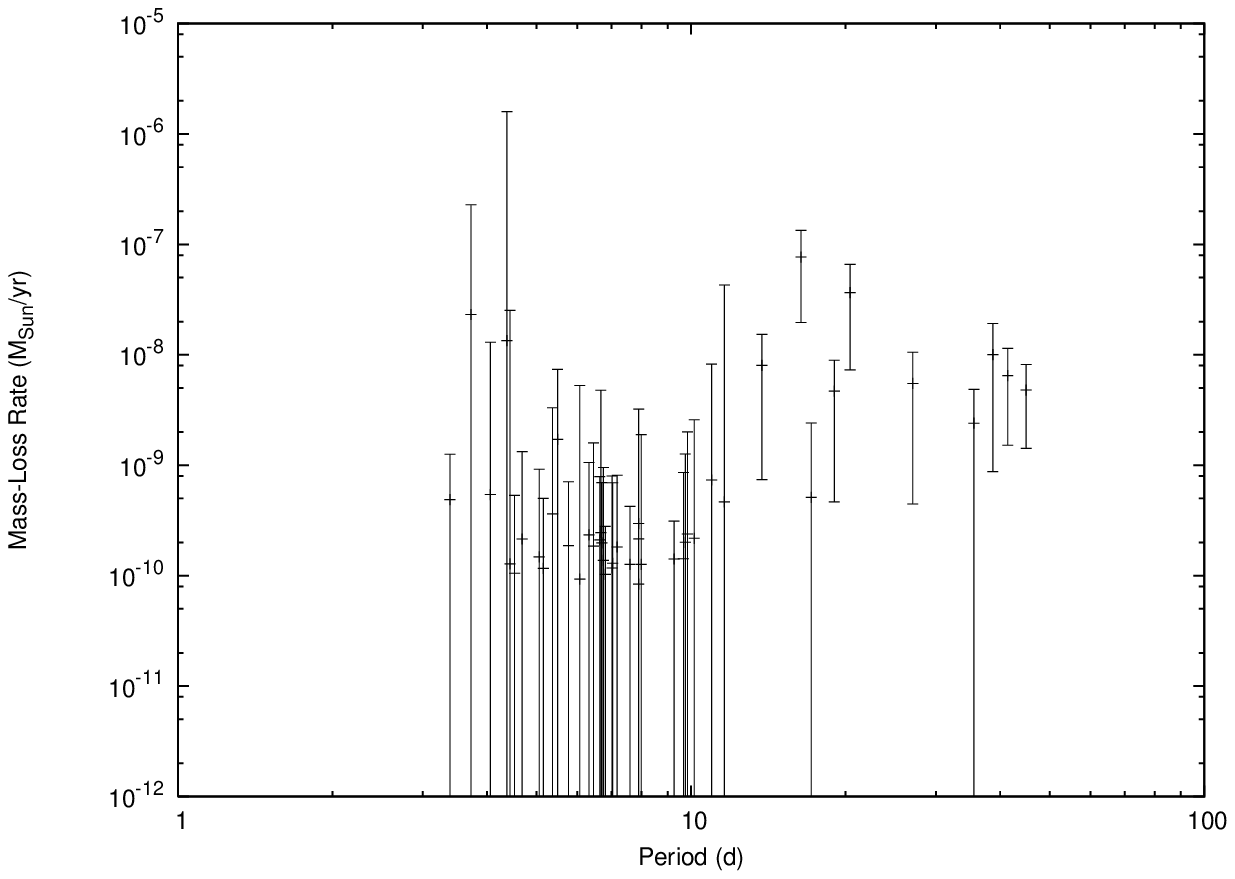}
 \caption{(Left) The fractional errors of the pulsation driven mass-loss rates for the sample of galactic Cepheids as a function of period. (Right) The mass--loss rates of the Cepheids as a function of period with errors calculated using Equation \ref{e49a} added.  The errors appear related to the enhancement of the mass loss due to pulsation and shocks and the plot shows that uncertainty of the mass--loss rate has a similar non--linear behavior as a function of mass as the mass--loss rate which is shown in Figure \ref{f12}. }
        \label{f12a}
     \end{center}
\end {figure}

In this work, the uncertainty of the luminosity is about $10\%$, the uncertainty of the mass is about $25\%$ to account for the error in the PMR relation, and the uncertainty of  the amplitude of variation of radius is about $5\%$ based on the uncertainty of the projection factor \citep{Nardetto2007}.  The uncertainty of the effective temperature is $\pm 200K$ while the uncertainty of the radius is calculated from the luminosity and effective temperature errors.  The fractional error of the pulsation driven mass--loss rate is shown in Figure \ref{f12a} (Left) as a function of period and  the mass--loss rates are plotted as a function of period with error bars in the right panel. This shows that the error of the mass loss is sensitive to the balance of gravity, and the uncertainty of the mass and luminosity has little effect unless they are at least an order of magnitude less.  This reiterates the importance of the balance of forces in determining the mass--loss rates.  While the uncertainties of the predicted values of the mass--loss rates are large in some cases, the observational determinations of mass--loss rates (as shown in the next section) also have large uncertainties.

\section{Comparison to Observed Mass--Loss Rates}
\begin{table*}[t]
\begin{center}
\begin{tabular}{lcl}
\hline
Name &$ \dot{M} (M_\odot /yr)$  & Reference\\
\hline
RS Pup & $3.5\times 10^{-6}$&\cite{Deasy1988} \\
$l$ Car &$ 2\times 10^{-8}$& \cite{Bohm-Vitense1994}\\
R TrA& $3\times 10^{-9}$&\cite{McAlary1986} \\
S Mus & $< 10^{-9}$ & \cite{Rodrigues1992} \\
$\delta$ Cep & $< 5.5\times 10^{-9}$ & \cite{Welch1988}\\
$\eta$ Aql & $< 6.1\times 10^{-9}$& \cite{Welch1988}\\
T Mon & $< 4.2\times 10^{-8}$& \cite{Welch1988}\\
\hline
\end{tabular}
\end{center}
\caption{The inferred mass--loss rates of various Cepheids from the literature.} 
\label{t4}
\end{table*}
There have been only a small number of measurements of mass--loss rates for Cepheids.  There are two main methods: observing the near and mid--infrared flux excess from dust and using that to determine mass loss, or from emission lines inferring large velocities and particle densities to measure mass loss.  The Cepheids with estimated mass--loss rates that are coincident with those used in this work are listed in Table \ref{t4}.  By comparing the measurements with the predicted mass--loss rates including pulsation and shocks, it is apparent the predictions are lower than the measured rates by about a factor of $10$, except for RS Pup which is a several orders of magnitude different.

 For the case of $l$ Car, \cite{Bohm-Vitense1994} used ultraviolet spectra to detect emission lines of C II, C IV, Mg II and O I. They argued carbon emission lines provide evidence for mass loss because they require velocities of order $100$ $km/s$ to form, which is the same order of magnitude as the escape velocity.  The Mg II lines are seen to have two emission components surrounding a broad central absorption profile that is relatively constant with pulsation phase. The width of  the central absorption profile of magnesium lines is believed to be a result of circumstellar material or interstellar material or both. \cite{Bohm-Vitense1994} use the velocities inferred from the lines to calculate an optical depth for the absorption shell and thus a column density.  The column density depends on the choice of the turbulent velocity, and the uncertainty in the turbulent velocity causes an exponential change in the estimate of the column density; for this reason \cite{Bohm-Vitense1994} argued the mass--loss rate was uncertain by two orders of magnitude.  Because \cite{Bohm-Vitense1994} used a value for the turbulent velocity near the lower limit, their mass--loss rate is an upper limit.  The mass loss derived by these authors depends also on the distance to $l$ Car, which was taken to be $d = 400$ $pc$. However, recent parallax measurements \citep{Benedict2007} suggest $l$ Car is about $500$ $pc$ away, thus lowering this upper limit  as well.  Given these qualifications, the value of the mass--loss rate predicted here, $2.4\times 10^{-8}M_\odot/yr$, is consistent with the result of \cite{Bohm-Vitense1994}.
 
IUE observations of S Mus \citep{Rodrigues1992} showed features in the absorption profile due to the Cepheid, its B5V companion and possibly the wind from the Cepheid.  The authors modeled the line profile by assuming a velocity law for the wind, which they assumed to be spherically symmetric and unperturbed by either pulsation or the companion.  Using a $\beta$--law for the wind, $v_{\infty}(1 - R_*/r)^\beta$, gives a mass--loss rate of approximately $1.1 - 1.3\times 10^{-10}M_\odot/yr$ by fitting the observed absorption profile.  Using different velocity laws, such as exponential and power--laws, had a dramatic effect on the mass--loss rate, with the maximum estimate being $2.5\times 10^5$ times larger than the minimum estimate.  The result from using the $\beta$--law alone is uncertain by a factor of about $6$ according to the authors, due to issues of fitting the continuum and the assumed abundances for the system.  Therefore the result is an optimistic upper limit. The important conclusion is that our predicted mass loss is small like the observed value, though both may be wrong if the lower limits of the mass determined by \cite{Evans2006} is the true value of the mass as shown in Figure \ref{f12}.  The mass--loss rate will still be consistent with that calculated by \cite{Rodrigues1992}.

\cite{McAlary1986} used IRAS observations to detect Cepheids and found modest infrared excess in several. By assuming the infrared excess is due to dust that formed in a stellar wind, the authors estimated mass--loss rates of the order $10^{-9}-10^{-8}M_\odot/yr$ and worked out the case of R TrA.   The calculated mass--loss rate for R TrA is $4\times 10^{-9}M_\odot/yr$, about a factor of 9 greater than that found here.  One of the differences in the mass--loss calculation may be due to differences in the choice of stellar radius and temperature,  \cite{McAlary1986} used a larger radius and smaller temperature.  A larger radius will affect the calculated density of the shell while the smaller temperature will predict a larger flux excess.  Therefore the two estimates for the mass--loss rate of R TrA are probably consistent.

Analysis of IRAS observations was repeated by \cite{Deasy1988} who found mass--loss rates consistent with the previous work.  Furthermore \cite{Deasy1988} derived a mass--loss rate from the geometric structure of the surrounding nebula for RS Pup.  The observations of \cite{Havlen1972} showed the reflection nebula was associated with the Cepheid and modeled the nebula as a set of four concentric dust shells, each having a mass of $0.05$--$0.1M_\odot$.  By assuming a velocity for the stellar wind, it is possible to determine a timescale for the shells, but the timescale of the innermost shell is much longer than for the others. This implies the existence of an undetected shell based on the argument the shell timescales are related to the crossing time of a Cepheid on the instability strip. The mass--loss rate is approximated by the dust mass of the shell and the lifetime of the crossing of the instability strip. By assuming a shell mass based on the spacing of shells, the mass of the innermost shell may be overestimated, especially as it has not been detected and the shell is still being generated. Therefore the estimate is a measure of the upper limit of the mass--loss rate.

\cite{Welch1988} observed a sample of Cepheids at radio wavelengths to measure mass--loss rates.  They modeled an ionized wind at temperatures of order $10^4$--$10^5$ K. The mass--loss rate is not sensitive to this assumption, but the temperature should not exceed $10^4$ K, the temperature where shocks would be generated in a Cepheid.  The authors also assume a wind velocity of order $100$ $km/s$, which is not necessarily true. The value is chosen because the observed components of the wind must exist at a radius greater than $4R_*$, meaning a lower wind velocity can be used if the components are older and have moved a further distance from the star.  There is also systematic uncertainty about the choice of distances to these Cepheids; parallax measurements indicate a difference of order $20$--$30\%$ \citep{Feast1997,Benedict2002, Benedict2007}.  The mass--loss rates are consistent, although the model devised by \cite{Welch1988} is different than the one used here.
   
The main conclusion to be drawn from these observed mass--loss rates is they are upper limits, with large uncertainties.  These upper limits are orders of magnitude larger than the predicted mass--loss rates derived by assuming only radiative--driving, implying there must be additional driving forces, lending credence to the pulsation $+$ shock mechanism described in this work.

\section{Model Infrared Excess}
The works of \cite{McAlary1986} and \cite{Deasy1988} both discuss the infrared flux excess of Cepheids in the IRAS bands. \cite{Kervella2006} and \cite{Merand2006,Merand2007} discovered excess K--band flux using interferometry, and \cite{Evans2007} discuss using the Spitzer Space Telescope to search for IR excess.  Predictions of infrared excess provide a useful test of the mechanism for mass loss in Cepheids.

Infrared excess can be produced in two distinct ways: by hot ionized winds or by cool dusty winds.  Since the wind model in this work assumes it is driven by mechanical energy, the wind is not hot so a dusty wind is presumed.  The wind is in radiative equilibrium with the Cepheid, so the temperature is dependent on the stellar temperature and is inversely proportional to the square of the distance from the star.  Dust forms in the wind when the temperature is near $1500$ K, the condensation temperature for dust.   Following the discussion from \cite{Lamers1999} for the luminosity of an optically thin dusty wind, one can represent the temperature structure as
\begin{equation}\label{e50}
T(r) = T_{\rm{eff}} \left(\frac{2r}{R_*}\right)^{-2/5}.
\end{equation} 
By rearranging, one can determine the condensation radius: for  $\delta$ Cep this is about $14.7R_*$, while for the coolest Cepheid in the sample, SV Vul, the condensation radius is about $7.9R_*$. 

  To calculate the dust luminosity, it is assumed that the dust is forming far enough from the star that pulsation and shocks will not affect the structure.  From the above calculation, however, this is not an ideal assumption, at least in the inner edge of the dust shell.  The total luminosity of the dust at a given frequency is derived by \cite{Lamers1999},
\begin{equation}\label{e51}
 L_\nu = \frac{3}{4\pi}\frac{<a^2>}{<a^3>}\frac{1}{\bar{\rho}_d} \frac{\dot{M}_d}{v_d}Q_\nu^A\int_{r_{\rm{min}}}^{r_{\rm{max}}} B_\nu(T_d)  \left\{1 - \frac{1}{2}\left(1 - \sqrt{1 - (R_*/r)^2}\right)\right\}dr.
\end{equation}
where the quantities with a subscript $d$ refer to dust. The terms in this expression are evaluated in the following way.  The dust is assumed to be graphite with $\rho_d = 2.2 g/cm^3$.  The mass--loss rate of dust is one--hundredth of the total mass--loss rate, based on the galactic gas--to--dust mass ratio.  Inside the condensation radius the value of $\dot{M}_d$ is zero.  The velocity, $v_d$, is the velocity of the wind at the distances from the star being considered. The value of $v_d$ that is used is the mean terminal velocity of the wind averaged over one period of pulsation.  To simplify the integration, $r_{\rm{min}} = R_*$ and $r_{\rm{max}} = \infty$;  if the velocity is $100$ $km/s$ then the wind would travel a distance of order $100$ $pc$ in one million years. The dust will contribute to the infrared excess only at a much smaller distance, so the assumption is reasonable. The dust is assumed to have a grain size distribution as given by \cite{Mathis1977}, where the number density is $n(a)da \propto Ka^{-3.5}da$.  If the grain size ranges from $0.005$ $\mu m$ to $0.25$ $\mu m$, then the term $<a^2>/<a^3>$ is $ \approx 40$ $ \mu m^{-1}$.  The absorption efficiency, $Q_\nu^A$, is based on the argument that most of the dust absorption will be at optical wavelengths and will be of order $2$ \citep{Jones1976}. Using these values, Equation \ref{e51} can be used to predict the luminosity of the dust shell in the infrared wavelengths of the VLTI, CHARA, IRAS and Spitzer.
\begin{figure}[t]
      \begin{center}
\epsscale{0.75}
    \plotone{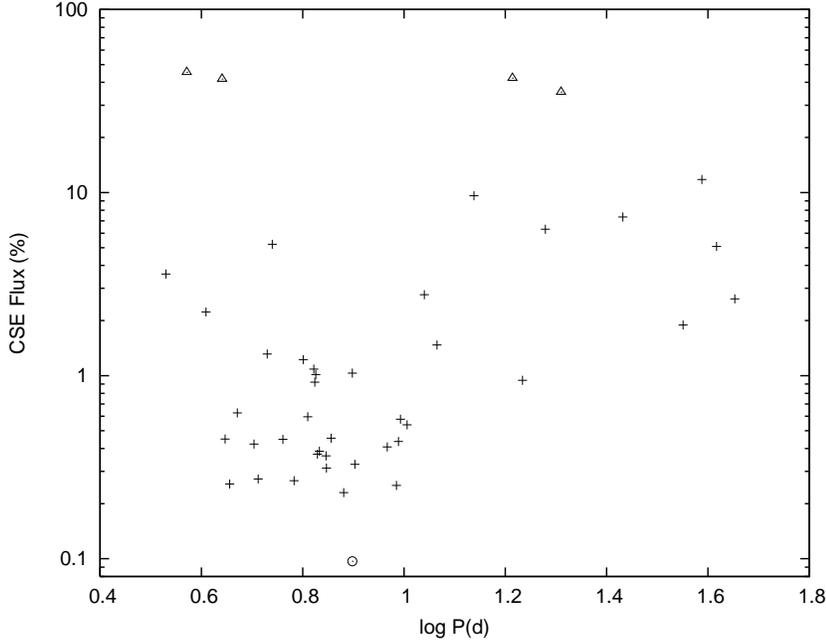}
 \caption{The predicted fraction of the total luminosity of a Cepheid contributed by the presence of dust at $2.2$ $\mu m$. The circle represents the example of the smallest predicted circumstellar emission (CSE) flux, RX Cam, and the triangles represent RT Aur, V Vel, X Cyg and RZ Vel which have the largest predict CSE fluxes of the sample.}
        \label{f5a}
     \end{center}
\end {figure}  
\begin{table}[t]
\begin{center}
\begin{tabular}{lcc}
\hline
Name &Observed CSE $\%$ & Predicted CSE $\% $\\
\hline
$\delta$ Cep & $1.5\pm 0.4$ & $1.3$\\
$l$ Car & $4.2\pm 0.2$ & $1.9$\\
Y Oph & $5.0\pm 2.0$&$0.95$\\
\hline
\end{tabular}
\end{center}
\caption{The flux of circumstellar shells of Cepheids relative to the total flux observed using K--band interferometry compared to the predicted flux of circumstellar shells.} 
\label{t4a}
\end{table}
\begin{figure}[t]
      \begin{center}
	\epsscale{0.48}
     \plotone{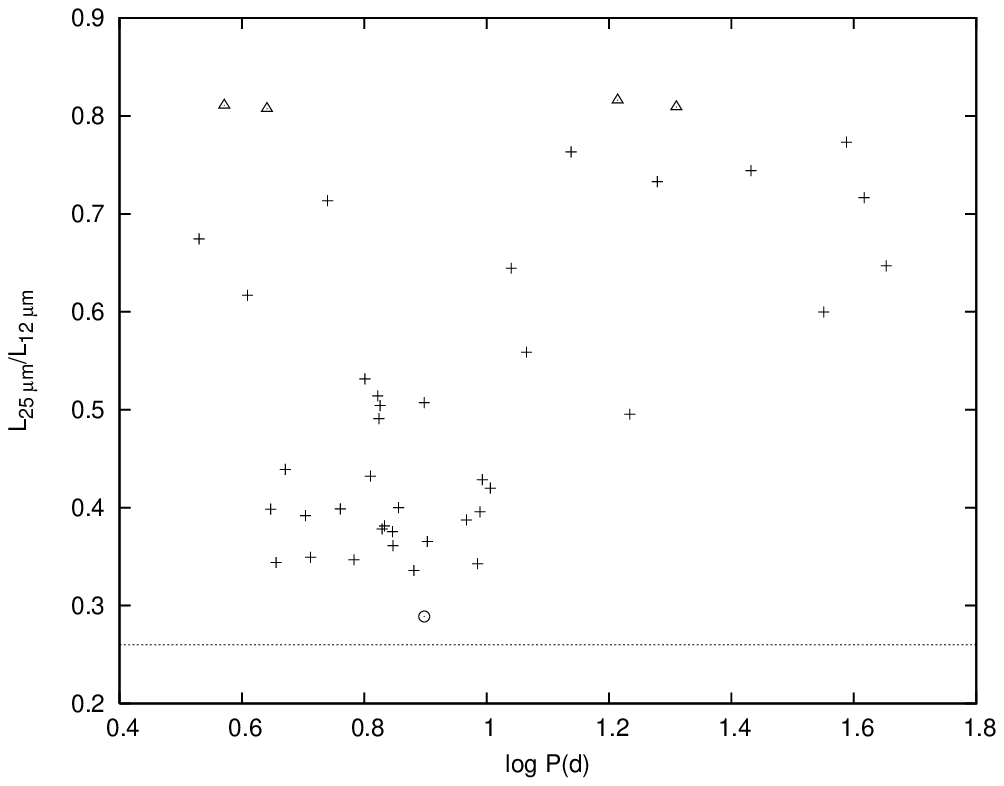}
     \plotone{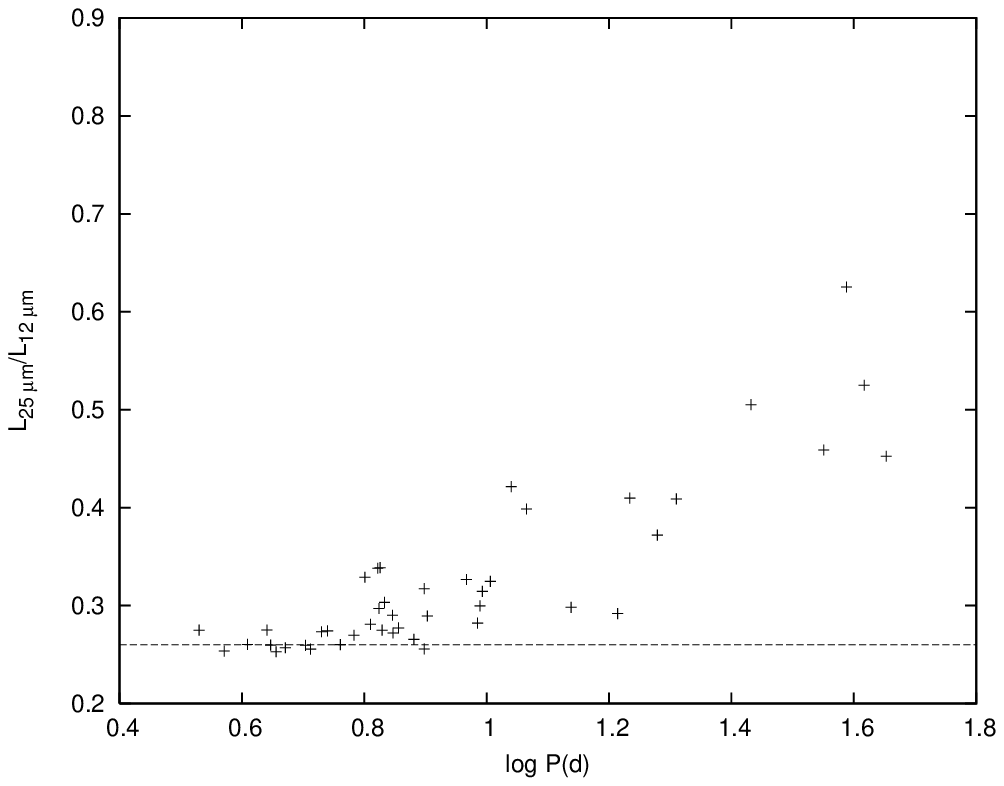}
 \caption{(Left Panel) The ratio of the luminosity of the wind at $25$ $\mu m$ and $12$ $ \mu m$ based on the predicted pulsation mass--loss rates.  The ratio of the luminosity at these wavelengths is about $0.26$ when there is no infrared excess, represented by the dashed line. The Cepheid, RX Cam, is represented by the circle at $\log P \approx 0.9$ with the smallest infrared excess of the sample. The triangles represent the largest predicted flux excess for the Cepheids for RT Aur, V Vel, X Cyg and RZ Vel. (Right Panel) The ratio of the luminosity of the wind at $25$ $\mu m$ and $12$ $ \mu m$ for the predicted mass--loss rates based on radiative driving alone.  The ratio of the luminosity at these wavelengths is about $0.26$ when there is no infrared excess, represented by the dashed line. }
        \label{f6}
     \end{center}
\end {figure}  
\begin{figure}[t]
      \begin{center}
	\epsscale{0.3}
 	\plotone{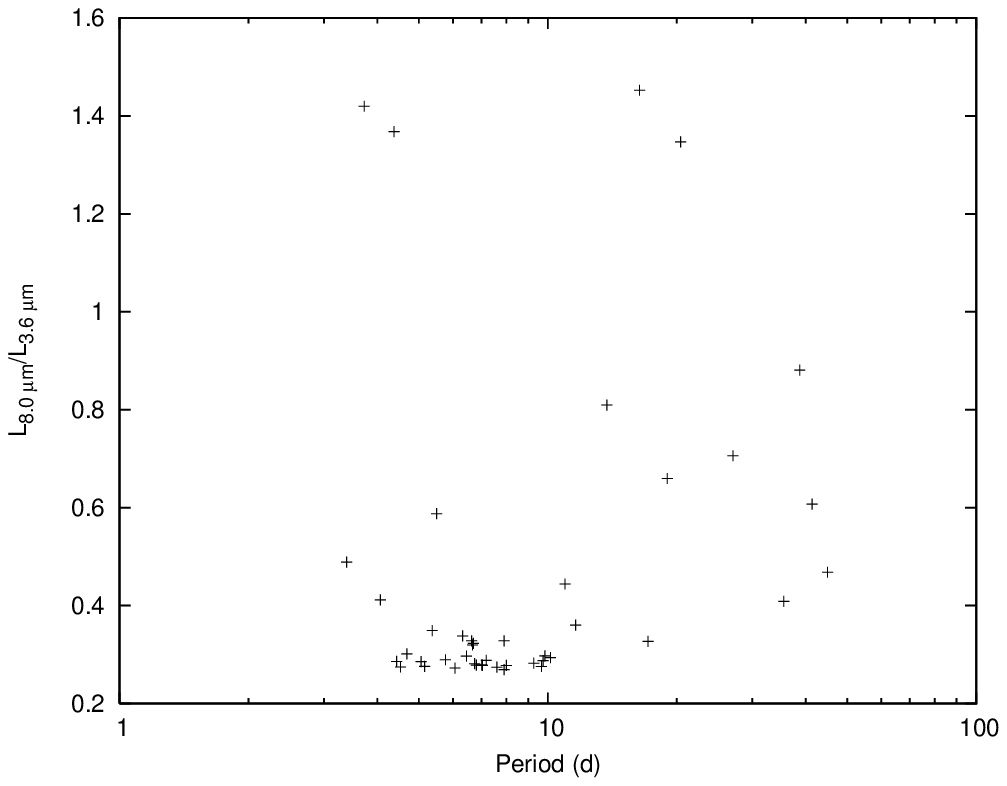}
	\plotone{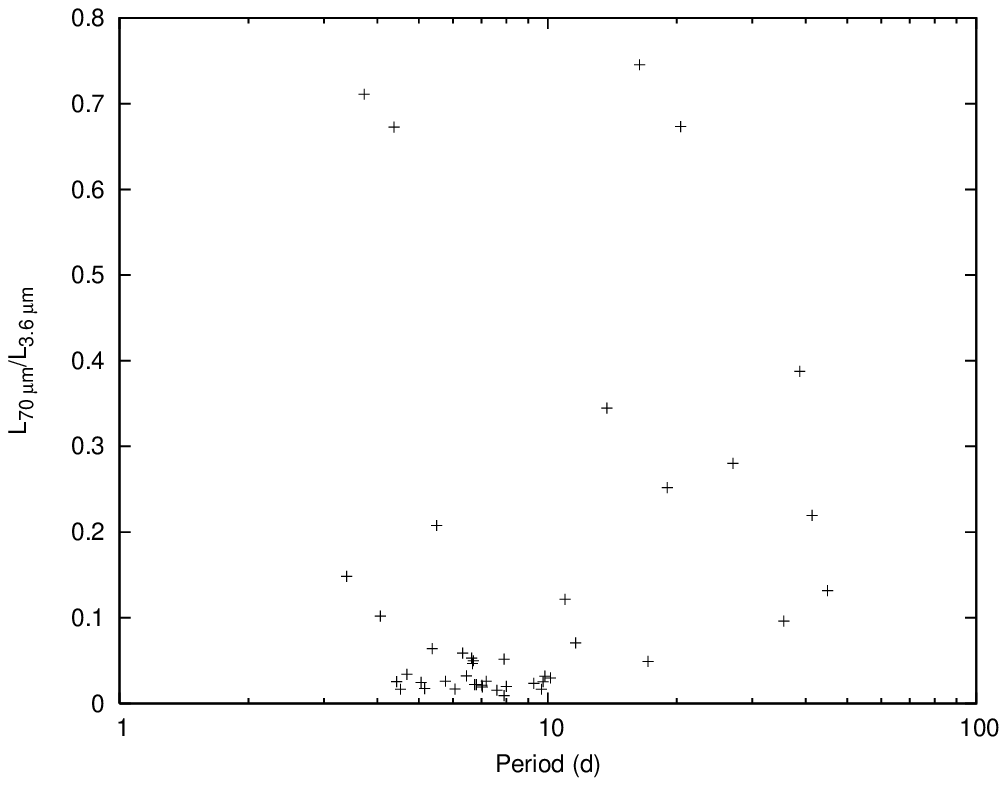}
	\plotone{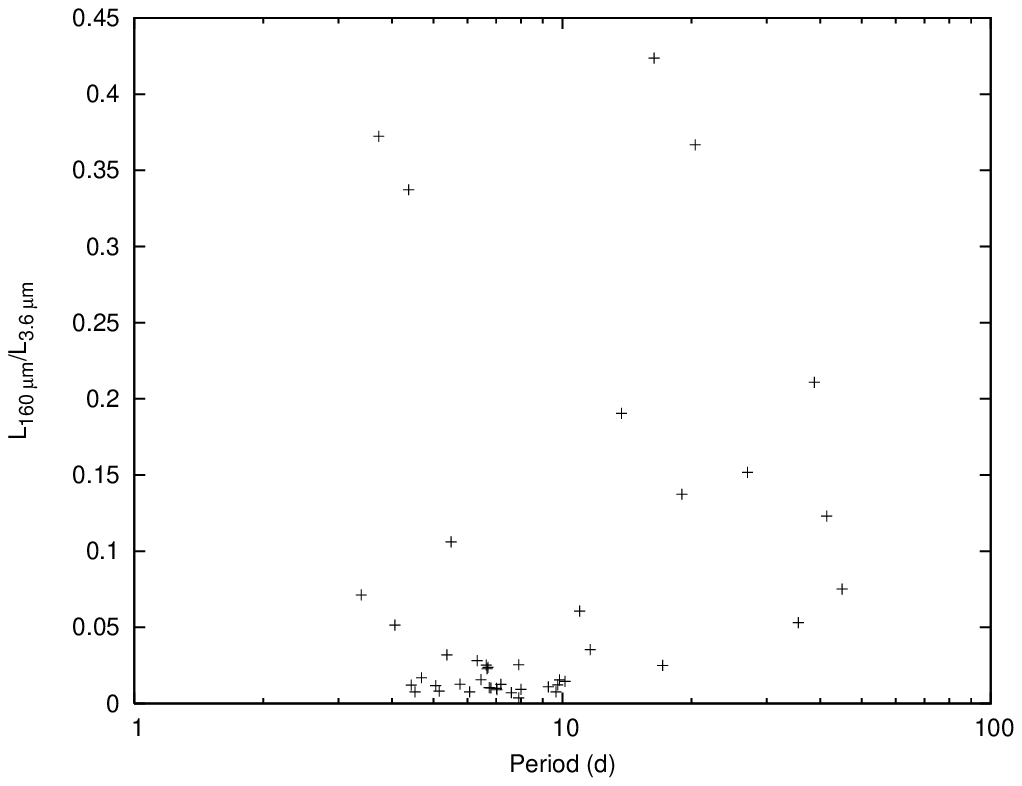}
 \caption{The ratio of the predicted luminosity at  $8.0$ $\mu m$ (Left Panel), $70 $ $\mu m$ (Middle Panel), and $160$ $\mu m$ (Right Panel) relative to the predicted $3.6$ $\mu m$ luminosity of the sample of Cepheids. }
        \label{f7}
     \end{center}
\end {figure} 

The infrared excess found from interferometric observations of Cepheids is summarized by \cite{Merand2007}, where the authors list the fraction of the total flux contributed by circumstellar shells, defined as the circumstellar emission (CSE) flux. The results for the fundamental pulsating Cepheids are shown in Table \ref{t4a}. The non--pulsating supergiant $\alpha$ Per was observed as well and was found to have almost no contribution to the $2.2$ $\mu m$ flux due to circumstellar material.  The infrared excess due to the predicted mass loss for the sample of Cepheids at $2.2$ $\mu m$ is shown in Figure \ref{f5a}, and the predicted CSE fluxes for the Cepheids observed with interferometry are given in Table \ref{t4a} for comparison.  These predictions are of the same order of magnitude as the observed, and differences may be due to the parameters in the dust model.  The CSE flux is linearly dependent on the ratio $<a^2>/<a^3>$, the mass--loss rate of dust, terminal velocity, grain density, etc.; a small change of any of these parameters will change the prediction.  The predicted CSE flux is most different for the cases of Y Oph and $l$ Car, which could be related to the uncertainty of the inner radius of the dust shell or to the predicted mass--loss rates being underestimated for these two Cepheids; an increase of the mass--loss rate by a factor of $4$ and $2$ respectively would match the observations.  In the enhanced CAK method this would require a decrease of chosen stellar mass of order of $1$ to $1.5M_\odot$  or an increase of stellar radius of order $7$--$15R_\odot$, or some combination of the two, both of which are within the range of observational uncertainty.  All things considered, the dust model is a consistent fit to the observations and the difference between the predicted and observed CSE is small. The predicted CSE fluxes of the sample Cepheids span a significant range, $0\%$ to $50\%$, due to the combination of the fundamental parameters.  This prediction can be tested.

The IRAS observations provide fluxes in its four bands for a large sample of Cepheids. \cite{Deasy1988}  found the ratio of $F(25$ $\mu m)/F(12$ $\mu m)$ ranges from roughly $0.25$ to $0.8$. The lower end is the expected ratio from just blackbody radiation with no excess radiation (Figure 1 of that paper).  The ratio of the infrared excesses from the predicted mass--loss rates is given in Figure \ref{f6} (Left).  It is striking that the ratio of fluxes exhibit the same range of values as in \cite{Deasy1988}. This agreement  is encouraging as the luminosity of the dust at these two wavelengths is much less sensitive to the dust parameters, and the contributing dust is much farther from the inner boundary of the dust shell. The infrared excess predicted for just radiative driven mass loss is shown in Figure \ref{f6} (Right), where the maximum excess is significantly less than that predicted for pulsation plus shock driven mass loss, and this prediction does not agree with the range of infrared excesses found by \cite{Deasy1988}.  This is further evidence that mass loss in Cepheids is driven by pulsation and not just radiation.   Note there appears to be a minimum in the flux ratio in Figure \ref{f6} (Left) near $\log P \approx  0.9$, which corresponds to $\log L/L_{\odot} \approx 3.5$;  this reflects the lack of mass--loss enhancement due to pulsation at that luminosity, which is consistent with the behavior as shown in Figure \ref{f4}.

Figures \ref{f5a} and \ref{f6} (Left) show that the mass--loss rates and dust model can reproduce the existing interferometric and IRAS observations, but the Spitzer Space Telescope can now provide new results.  Therefore it is useful to predict the infrared excess at wavelengths observed by Spitzer.  The luminosity of the Cepheids with dust shells are plotted in Figure \ref{f7}, relative to the $3.6$ $\mu m$ luminosity, at wavelengths $8, 70$, and $160$ $ \mu m$, respectively. The combination of the four wavelengths provide a test of mass--loss enhancement.  At $8$ $\mu m$, the excess is due to mass loss being enhanced by both pulsation and shocks and radiative driving at longer periods.   This would provide a test for the period dependence of mass loss.  Observations at the longer wavelengths provide a measure of smaller mass--loss rates because fluxes at these wavelengths are in the tail end of the stellar blackbody function as well as a measure of the size of the circumstellar shell.

There are extreme values of the predicted flux excess; from Figures \ref{f6} (Left) and \ref{f7} it is clear one Cepheid, at period $\log P = 0.9$, has minimal flux excess shown as a circle in the figures.  This is the model for RX Cam, where the predicted mass--loss rate is $8.4 \times 10^{-11}$, which is most likely due to the estimate of the mass $M = 10.8M_\odot$.  The infrared excess and mass--loss rate may easily be underestimated, but the result does show it is possible for Cepheids to have a low infrared excess similar to non--pulsating yellow supergiants. On the other hand, there are a number of Cepheids predicted to have very large IR excess, where the dust luminosity contributes most of the total luminosity at that wavelength. From Figure \ref{f6} (Left), there are four Cepheids with a CSE flux of order $50\%$: the short--period Cepheids RT Aur and V Vel and the 10--20 day period Cepheids X Cyg and RZ Vel.  The two short period Cepheids have a large enhancement of mass loss, both by almost $10^3$ times and hence the infrared excess is large.  The same is true for the longer period Cepheids but to a lesser extent. It is unlikely the excesses are overestimated, if it is assumed the mass--loss rates are reasonable.  Both V Vel and RT Aur have large effective temperatures for Cepheids, meaning the condensation radius is significantly large that it is not affected by the pulsation and shocks, supporting the predicted results.  In all,  the dust model discussed implies the measured infrared excesses and CSE fluxes can have a very large range, making it more challenging to use near--IR interferometry for distance estimates. 

\section{Mass Loss and Period Change}
The rate of period change of a Cepheid is a measure of its evolution on the instability strip due to the rate of change of effective temperature and luminosity  or equivalently the change of radius and luminosity.  The period change of Cepheids has been investigated by \cite{Turner2006} where the period change is
\begin{equation}\label{e52}
\frac{\dot{P}}{P} = \frac{6}{7}\frac{\dot{L}_*}{L_*} - \frac{24}{7}\frac{\dot{T}_{\rm{eff}}}{T_{\rm{eff}}}
\end{equation}
based on the Period--Mean Density relation and the small period dependence of the pulsation constant $Q \propto P^{1/8}$ \citep{Fernie1967}.  This relation assumes the mass of the Cepheid is constant with respect to time.  Starting with the Period--Mean Density relation, the period change can be re--derived to include mass loss,
\begin{equation}\label{e53}
\frac{7}{8}\frac{\dot{P}}{P} + \frac{1}{2}\frac{\dot{M}_*}{M_*} -\frac{3}{2}\frac{\dot{R}_*}{R_*} = 0.
\end{equation}
Substituting the luminosity and effective temperature for the radius for comparison to the result of \cite{Turner2006}, this relation becomes
\begin{equation}\label{e54}
\frac{\dot{P}}{P} = -\frac{4}{7}\frac{\dot{M}_*}{M_*} + \frac{6}{7}\frac{\dot{L}_*}{L_*} - \frac{24}{7}\frac{\dot{T}_{\rm{eff}}}{T_{\rm{eff}}}.
\end{equation}
This new relation implies the change of period is due to both evolution and mass loss.  Because the change of mass of a Cepheid is negative, the mass loss will always act to increase the period of pulsation.  

\begin{figure}[t]
      \begin{center}
	\epsscale{0.75}
 	\plotone{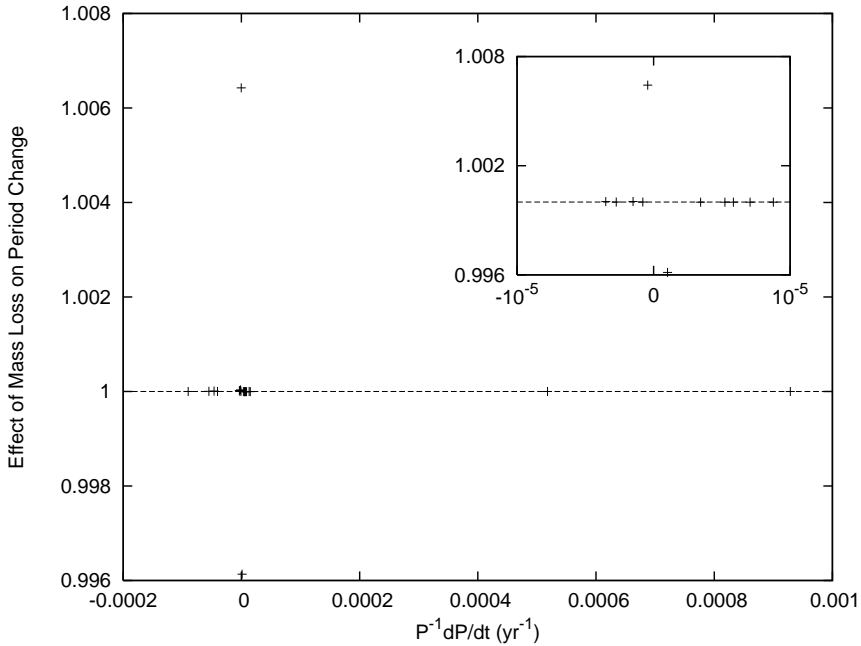}
 \caption{The fractional contribution of mass loss towards the period change for Cepheids as a function of period change.  Cepheids with a period change not affected by the mass--loss rate would fall on the dashed line.  Deviations from the dashed line measure how much mass loss plays a role. }
        \label{f10a}
     \end{center}
\end {figure}

The comparison of the observed and theoretical period change of Cepheids by \cite{Turner2006} showed models provide a reasonable fit to the observations overall.  There are, however, exceptions; one is the large number of Cepheids with a negative period change that is smaller in magnitude than that predicted by the models. This difference also appears in the comparison of positive period change but the  difference is smaller.  It is likely the models overestimate the rate of evolution, but a lower absolute rate of period change on the second crossing of the instability strip could be related to mass loss, which would increase the period change.  The role of mass loss on period change can be tested by plotting the fractional difference $(\dot{P}/P - 4\dot{M}/7M)/(\dot{P}/P)$ as a function of the fractional period change. Observed values of the period change for some of the sample of Cepheids are given in Table \ref{t5} taken from the literature \citep{Fernie1993, Turner1999,Berdnikov2000, Turner2004, Turner2005} and the comparison is shown in Figure \ref{f10a}.  Mass loss is a contributing factor to those Cepheids where the fractional difference is different from unity.   From Figure \ref{f10a}, there are only two Cepheids where the mass--loss rate would affect the period change of order $1\%$, X Cyg and RT Aur; both of which have large mass--loss rates and small rates of period change.  For the case of RT Aur, it should be noted the rate of period change was recently determined to be $\dot{P} = 0.082$ $s/yr$ implying the Cepheid is on its third crossing of the instability strip \citep{Turner2007}.  This new result differs from the rate of period change used here, $\dot{P} = -0.14$ $s/yr$,  but the magnitude of period change is still small, and mass loss will still significantly contribute to the rate of period change.  It can be inferred that mass loss does not play a significant role in changing their periods, complementing the result of \cite{Turner2006}.
\begin{table}[t]
\begin{center}
\begin{tabular}{lclc}
\hline
Name & $\dot{P} (s/yr)$ & Name & $\dot{P}(s/yr)$ \\
\hline

U Aql & $4.29$ & Y Oph & $766.82$ \\
$\eta$ Aql & $3.24$& BF Oph & $-1.23$ \\
RT Aur & $-0.14$ & V350 Sgr & $1.53$\\
VY Car & $-75.92$ &U Sgr & $5.11$\\
$\delta$ Cep & $-0.7$ & W Sgr & $3.84$\\
X Cyg & $1.45$ & X Sgr & $7.96$ \\
$\beta$ Dor & $12.83$& R TrA & $0.23$ \\
$\zeta$ Gem & $-78.88$& T Vul & $1.05$ \\
W Gem &  $-27.66$& SV Vul & $-214.3$ \\
T Mon & $2169.72$ & & \\
\hline
\end{tabular}
\end{center}
\caption{The rate of period change for some Cepheids with modeled mass loss.}
\label{t5}
\end{table}

The large mass--loss rates correlate to the minimal absolute period change of Cepheids.  
\begin{figure}[t]
      \begin{center}
	\epsscale{0.75}
 \plotone{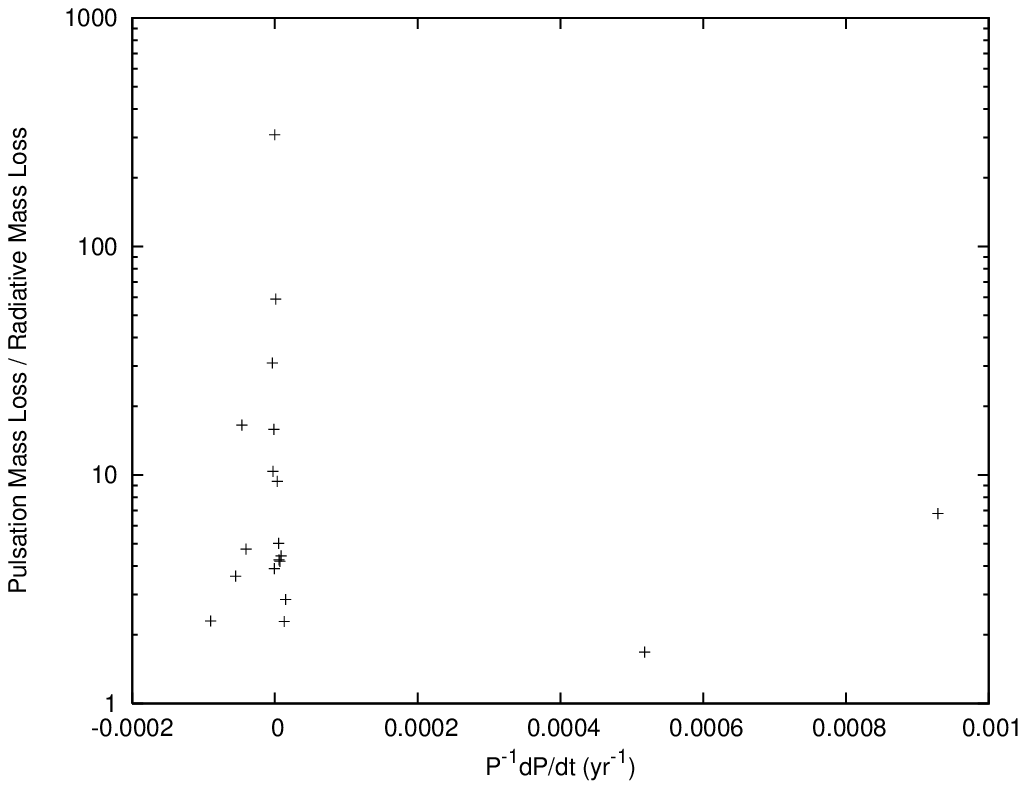}
 \caption{The enhancement of the mass loss, the ratio of the shock/pulsation mass loss and the mass loss from radiative driving, as a function of the period change $\dot{P}/P$. }
        \label{f11}
     \end{center}   
\end{figure}  
This point is emphasized in Figure \ref{f11}, where the enhancement of the mass loss by shocks plus pulsation is plotted against $\dot{P}/P$.  This implies mass loss is most enhanced when the Cepheid is evolving slowly through the instability strip.   This correlation can be produced in several ways. For example, when the mass--loss rate is large one might expect the radius of the Cepheid to be large, causing the gravity to be lower. A larger radius would also correspond to a larger luminosity, and  a larger luminosity would mean a lower rate of period change according to Equation \ref{e54}.  Another way to get a small period change is for the Cepheid to be at the blue edge of the instability strip where both the temperature and luminosity are largest for that crossing, meaning the ratios $\dot{T}_{\rm{eff}}/T_{\rm{eff}}$ and $\dot{L}_*/L_*$ are small.  At that point in the instability strip, the period is smallest, increasing the acceleration of gas due to both shocks and pulsation, which could increase the mass--loss rate.  The first explanation is seen in Figure \ref{f4} for the Cepheids with large mass--loss rates, but there are no examples of the second explanation.  The fact that there are few Cepheids from the sample evolving near the blue edge may account for this as well as the possibility mass loss may only be extreme very early upon entering the instability strip, making it very unlikely to catch one in the process.

The minimal role of mass loss upon the period change implies the period change is primarily due to evolution.  Near the blue edge of the instability strip the rate of evolution slows as the fractional change of luminosity and temperature is small.  Also near the blue edge, one might expect the mass--loss rate to be large because the period is lower, causing the acceleration due to pulsation and shocks to be larger.  For Cepheids on the second crossing, the rate of period change is $\dot{P} <0$, but near the blue edge of the crossing, mass loss will tend to decrease the absolute value of the rate of period change, and potentially even manage to change the sign of the rate of period change.  This can be tested by considering a sample of Cepheids where the period change is small and consistent with third crossing; the Cepheid X Cyg is likely on its third crossing with $\dot{P} = 1.52$ $s/yr$.  For X Cyg to be on its second crossing its mass--loss rate would need to be $\dot{M} \ge 2.1\times 10^{-5}M_\odot /yr$;  much too large, in the range observed for Wolf--Rayet stars. The Cepheid SX Car is another Cepheid consistent with being on the third crossing \citep{Turner2005}, with $\dot{P} = 0.07 $ $s/yr$, period of $4.86$ days and an inferred upper mass limit of  $5.7M_\odot$ \citep{Turner1996}.  This would require of minimum mass--loss rate of $\dot{M} = 1.67\times 10^{-6}M_\odot /yr$, which is an order of magnitude larger than the limit seen in Figure \ref{f2}. It has been shown that it is possible for mass loss to significantly affect the rate of period change at certain parts of the instability strip.

\section{Conclusions}
In this study, an analytic method is presented to describe the mass--loss rates of Cepheids using as input the global parameters that describe a Cepheid.  The derivation is based on the method for solving radiatively driven winds with additional energy supplied by shocks and pulsation.  It is assumed the wind is spherically symmetric, quasi--static and uses the Sobolev approximation.  The limits of the derivation are explored where in the simplest case the mass--loss rate is dependent on the ratio of the pulsation plus shocks and the effective gravity. It is shown mass loss can be enhanced by up to three orders of magnitude. 

The analytic method was tested using a set of Cepheids for which the necessary global parameters are given in the literature. For reference, the mass--loss rates, based on only radiative driving, were also calculated.  The energy added from pulsation and shocks increases the mass--loss rates, in some cases by orders of magnitude, predicting rates in closer agreement with those inferred from observations.  Furthermore the mass--loss behavior of Cepheids as a function of period is described, where the long period Cepheids have mass loss dominated by radiative driving but the short period Cepheids have a large range of mass--loss rates.  The large range of values implies the mass loss for short period Cepheids depend on the evolutionary state of the Cepheid.

The model provides reasonable estimates of mass--loss rates of Cepheids but there are uncertainties.    The mass is the largest uncertainty, as it is difficult to measure in general, so a Period--Mass--Radius relation is used.  The value of the radius also has an impact; the larger the radius, the lower the gravity and hence the larger the mass--loss rate but observed radii of Cepheids are precise with an uncertainty of a few percent \citep{Gieren1997}. There is uncertainty added due to the radiative driving coefficients $\alpha$ and $k$.  The mass--loss rate depends on the exponent $\alpha$, and variations of $\alpha$ can cause the predicted mass--loss rates to vary.  The value of $k$ is less important because it is small and $\dot{M} \propto k^{1/\alpha}$.  More detailed analysis of the mass loss of Cepheids would benefit from  calculations of $\alpha$ and $k$ at temperatures and gravities consistent with Cepheids over a finer grid.

 For some Cepheids, the mass loss seems to be driven primarily by the shocks generated in the atmosphere by pulsation.  The rates calculated are dependent on the accuracy of the pulsating atmosphere model by \cite{Fokin1996}.  Errors in the model of a Cepheid atmosphere with mass $M = 5.7M_\odot$, and $T_{\rm{eff}} = 5750$ K will affect other the mass--loss rates will be wrong due to the relative scaling.  It would be useful to verify the accuracy of the model of the shocks in pulsating atmospheres. Furthermore, models of pulsating atmospheres could be used to calculate the mass loss of the Cepheid directly to provide a consistency test of the analytic method.

The shock model may also be tested with observations.  Shocks propagating in a stellar atmosphere cause the layers in the star to move and this is reflected in the broadening of lines in a spectrum. \cite{Mathias2006} analyzed the time varying spectra of X Sgr and found evidence for two, possibly three, shocks moving through the atmosphere of the Cepheid in one pulsation period.  This result suggests spectroscopic observations of a Cepheid with intense phase coverage would provide information for understanding shocks in Cepheids.  X-ray and UV observations are important because shocks moving in the atmosphere will ionize material which could be detected as X-ray and UV variability \citep{Engle2006}.  The observations would provide a test for any hydrodynamic model of shocks in a Cepheid atmosphere.

It would be interesting to apply the theory of pulsation and shock induced mass loss to first overtone Cepheids and other radial pulsating variables.  As first overtone pulsation occurs at an earlier stage of pulsation than fundamental pulsation, analysis of mass loss in first overtone pulsating Cepheids would shed more light on the contribution of mass loss on the problem of the Cepheid mass discrepancy as well as contribute to the understanding of Cepheid evolution.  However, the shock model of \cite{Fokin1996}, that underlies this analysis, is valid only for fundamental pulsation and thus cannot be realistically applied to first overtone pulsation.  With an appropriate shock model, shock induced mass loss can be investigated with this theory.

The dust model proposed here provides a consistent match with existing observations.  In particular, the calculation of the ratio of the luminosity at wavelengths $25$ $\mu m$ and $12$ $\mu m$ exhibits the same behavior as shown by \cite{Deasy1988}.  The results, however, are dependent on the choice of dust parameters such as mean grain size.  New observations from Spitzer and at sub--millimeter wavelengths would provide constraints on the properties of the dust in the wind and on the mass--loss rates of Cepheids. Sub--mm observations would also provide information on the age of Cepheids.  Consider RS Pup with four observed shells, each associated with a particular crossing on the instability strip; a very strong sub--mm excess may indicate a large amount of cool dust in these shells.  Thus the amount of sub--mm excess may trace the mass--loss history of the Cepheid and its age.

The connection between mass loss and evolution has been explored for the observed Cepheids.  It was shown the period change of a Cepheid is dependent on the mass--loss rate at that time. However, for all the Cepheids with period changes quoted, the mass--loss rate provides an insignificant contribution, and, furthermore, the Cepheids with the largest calculated mass--loss rates have the slowest absolute value of period change.  This connection needs to be explored in greater detail with a larger set of observations.

Mass loss in the instability strip is an important concept to understand if one is to determine the physical structure and the cause of the mass discrepancy of Cepheids.  It is also important to constrain mass loss for observations in the near--IR and longer wavelengths if the period--luminosity relation is to be precise to a few percent, which is a motivation of interferometric observations.  The analytical model of mass loss of the Cepheids investigated here imply shocks and pulsation play a strong role driving the wind and increase the mass--loss rate by orders of magnitude.

\begin{acknowledgements}
HN would like to thank John Percy for careful reading and comments, and Tom Bolton for teaching the course on stellar winds that inspired this work.  The authors would also like to than the anonymous referee for helpful comments. The research was funded by the Walter John Helm OGSST and the Walter C. Sumner Memorial Fellowship.
\end{acknowledgements} 
\bibliography{wind_th}
\bibliographystyle{apj}

\end{document}